\documentclass[final,5p,times]{elsarticle}

\usepackage{amssymb}

\usepackage{graphicx}
\usepackage{subfigure}
\usepackage{multirow}
\usepackage{booktabs}
\usepackage{gensymb}
\usepackage{minibox}

\usepackage{eqparbox} 
\usepackage[x11names]{xcolor} 
\usepackage{color}

\journal{Computers \& Graphics}

\begin{document}

\begin{frontmatter}


\title{Epsilon-shapes: characterizing, detecting and thickening thin features in geometric models}


\author[CNR,CNR]{Daniela Cabiddu, Marco Attene}
\address[CNR]{CNR-IMATI Genova, Italy}

\author{}

\address{}

\begin{abstract}
We focus on the analysis of planar shapes and solid objects having thin features and propose a new mathematical model to characterize them. Based on our model, that we call an epsilon-shape, we show how thin parts can be effectively and efficiently detected by an algorithm, and propose a novel approach to thicken these features while leaving all the other parts of the shape unchanged. When compared with state-of-the-art solutions, our proposal proves to be particularly flexible, efficient and stable, and does not require any unintuitive parameter to fine-tune the process. Furthermore, our method is able to detect thin features both in the object and in its complement, thus providing a useful tool to detect thin cavities and narrow channels. We discuss the importance of this kind of analysis in the design of robust structures and in the creation of geometry to be fabricated with modern additive manufacturing technology.

\end{abstract}

\begin{keyword}


\end{keyword}

\end{frontmatter}


\section{Introduction}
\label{sec:intro}

Thickness information is important in a number of shape-related applications, including surface segmentation, shape retrieval and simulation. By aggregating points with a similar local thickness, Shapira and colleagues \cite{Shapira2008} developed a pose-invariant shape partitioning algorithm, whereas thickness was used in histogram form as a shape descriptor for retrieval purposes in \cite{Schmitt2015}. While considering shape thickness in general, a particular relevance is assumed by thin parts such as thin walls or wire-like features. These parts represent weak locations in physical structures, can be an obstacle for thermal dissipation, and can be subject to unwanted melting when traversed by intense electric currents. Furthermore, thin features are a serious issue in many manufacturing technologies. In plastic injection molding, for example, thin regions can restrict the flow of molten plastic \cite{Osswald:2008}, causing the mold to be only partially filled. In modern additive manufacturing technologies, too thin features may lead to small layers of the surface being peeled, thin shape parts being fully removed, and shape break-up in several parts due to narrow connections \cite{Telea:2011}.

The intuitive concept of ``shape thickness'' has been formalized in several ways in the literature, and diverse algorithms exist for its actual evaluation \cite{Shapira2008} \cite{kovacic:2010} \cite{Miklos:2010}. However, existing algorithms require a trade-off between accuracy and efficiency that can be hard to establish for demanding applications such as industrial 3D printing.

In this paper we introduce the concept of ``$\epsilon$-shape'' which is a mathematical model for objects having no features thinner than a threshold value $\epsilon$. Based on this concept, we describe an algorithm to exactly and efficiently detect thin features. Differently from existing methods, our approach focuses on thin features only, meaning that parts of the geometry which are thick enough are not subject to any complex analysis: this is the key to achieve efficiency without sacrificing the precision. Furthermore, our formulation allows to detect thin portions in both the model and its complementary part. Note that this characteristic is particularly important when analyzing 3D mechanisms \cite{Hegel:2015}: if parts which are supposed to move independently are separated by a too small space, they risk to be glued when printed.

To demonstrate the usefulness of our analysis in practice, we describe a novel algorithm to thicken those parts of the shape which are thinner than a given threshold value (e.g. the layer thickness for 3D printing applications) while leaving all the other parts unaltered.
 

\section{Related Works}
\label{sec:related}

The medial axis of a shape is the locus of points having more than one closest point on the shape’s boundary. The medial axis may be enriched by considering each of its points endowed with its minimum distance from the shape's boundary \cite{Choi1997}, leading to the so-called Medial Axis Transform (MAT). From a purely theoretical point of view, the MAT should be the reference tool to compute thickness information.

Unfortunately, as observed in \cite{Shapira2008}, computing the medial axis and the MAT of a surface mesh can be an expensive process, and the medial axis itself is hard to manipulate \cite{Amenta2001}. Furthermore, algorithms based on the MAT are too sensitive and tend to confuse surface noise with small features unless approximations are employed \cite{Miklos:2010} \cite{li:2015}. This motivated the emergence of the numerous alternative methods described in the remainder of this section.

\subsection{Voxel-based methods}
The analysis of medical 3D images has been an early calling application for thickness analysis. For Hildebrand and R{\"u}egsegger \cite{Hildebrand:1997}, the thickness related to a surface point (i.e. a skin voxel) is defined as the radius of the largest sphere centered inside the object and touching that point. This definition was turned into an algorithm in \cite{Dougherty2007} where, after having computed the discrete medial axis, a distance transform from it is employed to associate a thickness value to all the skin voxels. In a more efficient approach based on PDEs \cite{Yezzi:2003}, Yezzy and colleagues compute thickness as the minimum-length surface-to-surface path between pairs of surface points. Telea and Jalba \cite{Telea:2011} observe that the extension of Yezzi's method to higher-genus models is not evident, and propose an alternative algorithm based on the top-hat transform from multi-scale morphology \cite{Maragos1996}.

All these methods are suitable when the input comes in voxel form, but for objects represented in vector form (e.g. triangle meshes) an initial ``voxelization'' step is required which necessarily introduces a distortion. In principle, for 3D printing applications it is sufficient to keep this distortion below the printer’s resolution \cite{Telea:2011}. Even if this solution may work for low-resolution devices such as low-cost FDM-based printer's, it is too demanding for industrial printing where the layer thickness can be as small as 14 microns \cite{Stratasys}, which means that a $10^3$ centimeters cube would require more than 364 billion voxels.

\subsection{Mesh-based methods}
In \cite{Lambourne:2005} the concept of thickness at a point is defined based on spheres centered within the object and touching that point tangentially. Differently from the medial axis, these spheres may be not completely contained in the object. Though this method can be applied both on NURBS models and on piecewise-linear meshes, it has been shown to work well on relatively simple shapes only.

\begin{figure}
	\centering
	\subfigure[\label{fig:sdf_related_00}]{\includegraphics[height=1cm]{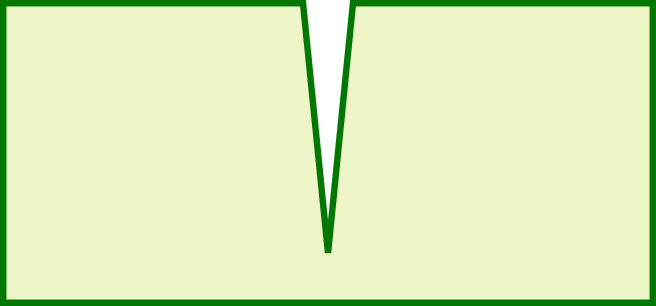}} \quad
	\subfigure[\label{fig:sdf_related_01}]{\includegraphics[height=2.5cm]{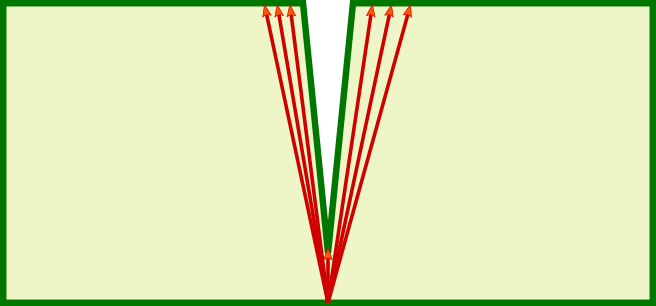}}
	\caption{SDF may return a result too far from the expectation. \subref{fig:sdf_related_00} Example of a shape with a narrow bottleneck in the middle. \subref{fig:sdf_related_01} Only a minority of casted rays touches the bottleneck tip. }
		\label{fig:sdf}
\end{figure}

The Shape Diameter Function (SDF) was introduced in \cite{Shapira2008} as a tool to perform mesh segmentation, and today is one of the most diffused methods to evaluate the thickness at a mesh point. The idea is to take a point on the surface and (1) compute the surface normal at that point, (2) cast a ray in the opposite direction, and (3) measure the distance of the first intersection of this ray with the mesh. Since this basic approach is too sensitive to small shape features and noise, Shapira and colleagues propose to use a number of casting directions within a cone around the estimated surface normal, and to keep a statistically relevant average of all the distances. Unfortunately this workaround may lead to completely unexpected results exactly in those cases that we consider to be critical. Imagine a tubular shape with a very narrow bottleneck on one side (see Fig. \ref{fig:sdf}). In this case only a minority of the rays hits the bottleneck tip, and thus the resulting average would be far from the expectation. This is even amplified in practice because, in an attempt to make the evaluation more accurate, the SDF method filters those diameters which are too far from the median value, which are considered to be outliers. Note that this is a serious issue in all those applications where potential structural weaknesses must be detected (e.g. industrial 3D printing). Though some improvements are possible to make the SDF calculation faster \cite{kovacic:2010} and less sensitive to noise \cite{Rolland:2013}, this intrinsic limitation remains.

\subsection{Thickening}
Automatic detection of thin parts is an important tool for a designer who is producing a shape model. If too thin parts are actually detected, the designer may take countermeasures such as thickening those parts. However, in contexts where the user is not expert enough (i.e. home users of consumer 3D printers), automatic thickening algorithms would help a lot. This assumption motivated the work by Wang and Chen \cite{Wang:2013} in the context of solid fabrication: models created by unexperienced designers may represent thin features by zero-thickness triangle strips, and \cite{Wang:2013} presents an algorithm to thicken these features so that they become solid and printable while keeping the outer surface geometry unaltered.

A generalization of morphological operators is used in \cite{Calderon:2014}, where thickening can be obtained by applying an overall dilation to the input.

The aforementioned methods do not adjust the amount of thickening depending on the local needs (i.e. the entire model is uniformly thickened). Conversely, based on an approximate medial axis, \cite{Stava:2012} cleverly analyses the geometry to predict where the printed prototype undergoes excessive stress due to manipulation and, besides adaptively thickening the weak parts, the method also adds little supporting structures to improve the overall object robustness. Note that while we focus on geometry only, in \cite{Stava:2012} the objective is to meet physical requirements, and some of the thin features may remain if they do not represent a weak part according to mechanical criteria. A comprehensive worst-case simulation is performed in \cite{Zhou:2013}, but in this case the algorithm proposed is limited to the analysis and does not include a thickening phase.

\subsection{Summary of contributions}
The method presented in this article strives to overcome the limitations of the existing approaches discussed so far. In particular, we provide an original well-defined mathematical model to represent thickness information in any dimension and, based on this model, define an efficient algorithm to detect thin features in 2D and 3D objects represented in piecewise-linear form. We show how the method can be used to detect thin features in both the object and its complementary part, and demonstrate how to speed up the calculation by limiting the complex analysis within thin areas, which are the subject of our investigation. Finally, we show how thickness information can be exploited to adaptively thickening the object only where necessary, while leaving all the other parts of the shape unchanged.


\section{Epsilon shapes}
\label{sec:definitions}



Let $S$ be a compact $n$-manifold with boundary, embedded in $\mathbb{R}^n$.
Let $x$ be a point lying on the boundary of $S$ and let $\epsilon$ be a real non-negative number ($\epsilon \in \mathbb{R}^+\cup\{0\}$).
Let $\Omega_{\epsilon}^x$ be a closed $n$-ball centered in $x$ and having radius equal to $\epsilon$.

We define two operators that we call \emph{$\epsilon$-sum} and \emph{$\epsilon$-difference}. 

An \textit{$\epsilon$-sum} $\sigma_{\epsilon}(S, x)$ is the result of the union of $S$ and $\Omega_{\epsilon}^x$

\[ \sigma_{\epsilon}(S, x) = S \cup \Omega_{\epsilon}^x \]

while an \textit{$\epsilon$-difference} $\delta_{\epsilon} (S, x)$ is the result of the difference between $S$ and the interior of $\Omega_{\epsilon}^x$

\[ \delta_{\epsilon}(S, x) = S - int(\Omega_{\epsilon}^x) \]

$\epsilon$-sums and $\epsilon$-differences are grouped under a common definition of \textit{$\epsilon$-modifications}. An $\epsilon$-modification of $S$ is \textit{topologically neutral} if the interior of its result is homeomorphic with the interior of $S$. It is worth noticing that the boundary of a topologically neutral $\epsilon$-modification may not be an $(n-1)$-manifold.

A shape $S$ is an \textit{$\epsilon$-shape} if all its $\epsilon'$-modifications are neutral, for each $\epsilon' \le \epsilon$. If $S$ is a $\epsilon$-shape, then $S$ is also an $\epsilon'$-shape for each $\epsilon' < \epsilon$. Let $S$ be an $\epsilon$-shape. $S$ is \textit{maximal} if no $\epsilon' > \epsilon$ exists so that $S$ is also an $\epsilon'$-shape. 

\begin{figure}[!h]
	\centering
	\includegraphics[width=0.15\textwidth]{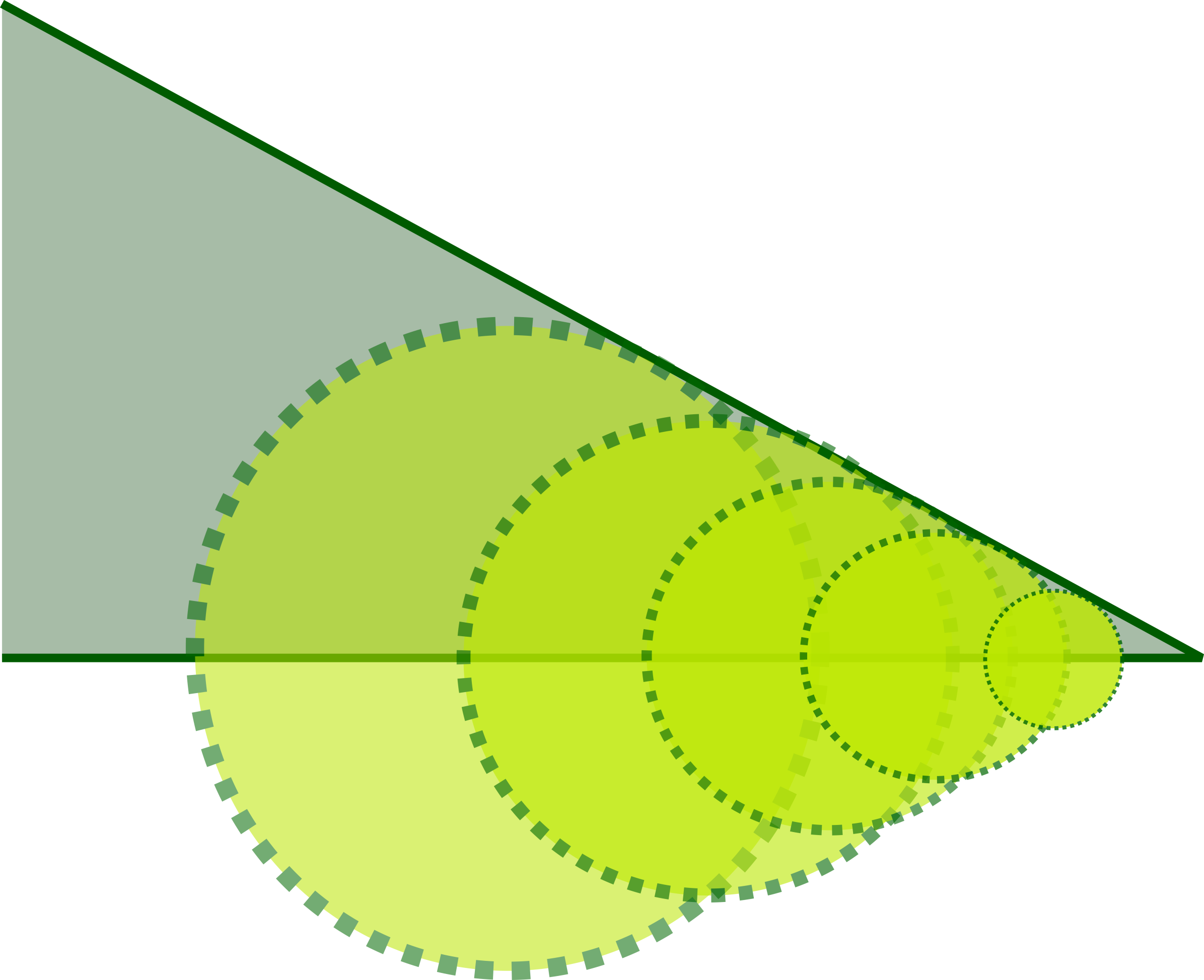}
	\caption{Acute angles. Any arbitrarily small disk induces a non-neutral $\epsilon$-difference if its center is sufficiently close to an acute angle.}
	\label{fig:acute_angle}
\end{figure}

Although any compact $n$-manifold is an $\epsilon$-shape for some $\epsilon \ge 0$, it is worth observing that there are some such manifolds for which
this is true only for $\epsilon = 0$. An example in 2D is represented by polygons with acute angles, where any arbitrarily small disk induces a non-neutral $\epsilon$-difference if its center is sufficiently close to the acute angle (Figure \ref{fig:acute_angle}). However, even in these cases, we can analyze the boundary in a
point-wise manner and construct what we call an $\epsilon$-map.

We say that an $\epsilon$-sum $\sigma_{\epsilon}(S,x)$ is \textit{strongly neutral} if all the $\sigma_{\epsilon'}(S, x)$ with $\epsilon' \le \epsilon$ are topologically neutral. An analogous concept is defined for $\epsilon$-differences. A positive $\epsilon$-map is a function $E^+: \partial S \rightarrow \mathbb{R}^+\cup\{0\}$ that maps each point $x$ of the boundary of $S$ to the maximum value of $\epsilon$ for which the $\epsilon$-sum at $x$ is strongly neutral. A negative $\epsilon$-map $E^-$ is defined as $E^+$ while replacing $\epsilon$-sums with $\epsilon$-differences. An $\epsilon$-map $E: \partial S \rightarrow \mathbb{R}^+\cup\{0\}$ is defined as $E(x) = min(E^+(x), E^-(x))$.
The minimum of $E(S)$ is the value $\epsilon$ for which $S$ is a maximal $\epsilon$-shape. We extend this concept by saying that, if $\epsilon$ is the minimum of $E^+(S)$, then $S$ is a \textit{positive} $\epsilon$-shape. If $\epsilon$ is the minimum of $E^-(S)$, then $S$ is a \textit{negative} $\epsilon$-shape.

A positive $\epsilon$-map represents our formalization of the intuitive concept of shape thickness, whereas a negative $\epsilon$-map describes the thickness of the shape's complement. The following Sections \ref{sec:2d} and \ref{sec:3d} describe an algorithm to compute these maps in 2D and 3D respectively.

\section{Planar shape analysis}
\label{sec:2d}

\begin{figure}[!h]
	\centering
	\includegraphics[width=0.2\textwidth]{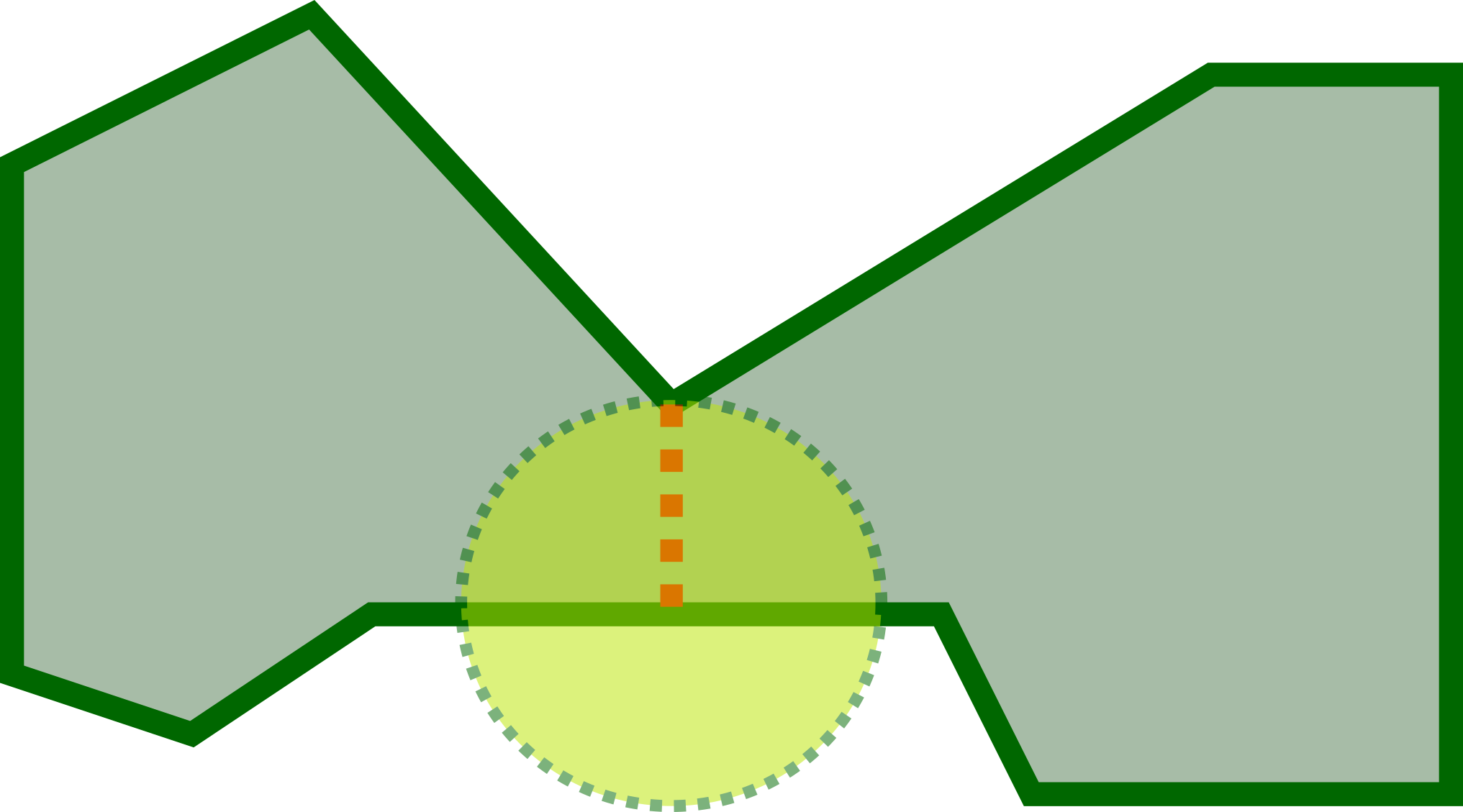}
	\caption{Example of a local minimum in the middle of an edge.}
	\label{fig:tangent}
\end{figure}

In our scenario the input is a single polygon $P$, possibly non-simply
connected. Hence the boundary of $P$, $\partial P$, is a 1-manifold made of a number of
edges connected to each other at vertices.
In a straightforward approach, the value of $E(v_i)$ may be
associated to each vertex $v_i$ of $P$.
However, we observe that a so-defined $\epsilon$-map is not sufficient
to represent all the thin features, since local minima of $E(P)$
may correspond to points lying in the middle of some edges (see Figure \ref{fig:tangent}).
To enable a comprehensive and conservative representation of all the
thin features, the aforementioned $\epsilon$-map must be encoded while
considering internal edge points as well.
If a local minimum happens to be on a point $x$ in the middle of an
edge, we split that edge at $x$ so that $E(x)$ is represented.

Our algorithm to compute $E(P)$ is based on a region growing
approach. Intuitively, we imagine to have an infinitely small disk
centered at a point $x$ on $\partial P$, and imagine to grow its radius
in a continuous manner. Initially, the portion of $\partial P$ contained
in the disk is made of a single open line, and we keep growing the
radius as long as the topology of this restricted $\partial P$ does not
change. It might happen, for example, that the line splits into separate
components, or that it becomes closed (e.g. when the disk becomes large
enough to contain the whole $P$). The maximum value reached by the radius
is the value of $E(x)$ that we eventually associate to $x$.
To turn this intuitive idea into a working algorithm, we first create a
constrained triangulation of $P$'s convex hull, where all the edges of $P$
are also edges of the triangulation. Then, we discretize the growing
process by calculating relevant ``events'' that may occur either when the
growing disk hits an edge of the triangulation, or when it hits one of
its vertices. This approach is close in spirit to Chen and Han's
algorithm \cite{Chen:1990} to calculate geodesic distances on polyhedra,
where the topology of the evolving front is guaranteed not to change
between two subsequent events.

To simplify the exposition, we first describe how to compute $E$
at a single vertex (Section \ref{sec:2d_vertex_method}), and then show how to extend the method
to find possible local minima of $E$ along an edge (Section \ref{sec:2d_edge_method}).
Finally, we describe an optimized procedure to compute the value of $E$ on
the whole polygon while possibly splitting its edges at local minima
(Section \ref{sec:2d_queue}).

%
%

\subsection{Thickness for a Single Vertex}
\label{sec:2d_vertex_method}
\begin{figure*}[!h]
	\centering
	
	\subfigure[\label{fig:proc1} Convex hull and triangulation.]
	{\includegraphics[width=4cm]{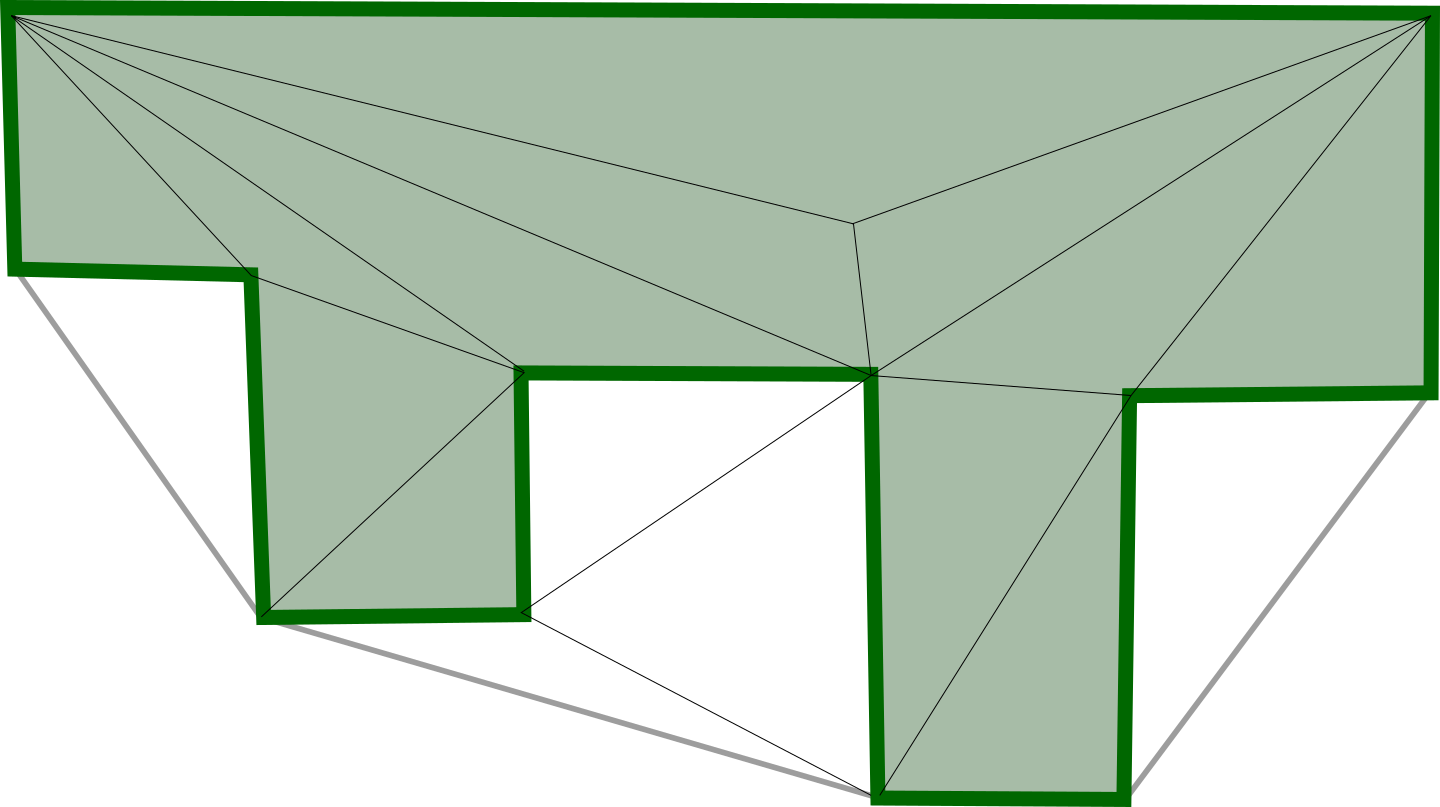}} \quad 
	\subfigure[\label{fig:proc2} $R(x)$ at the first iteration.]
	{\includegraphics[width=4cm]{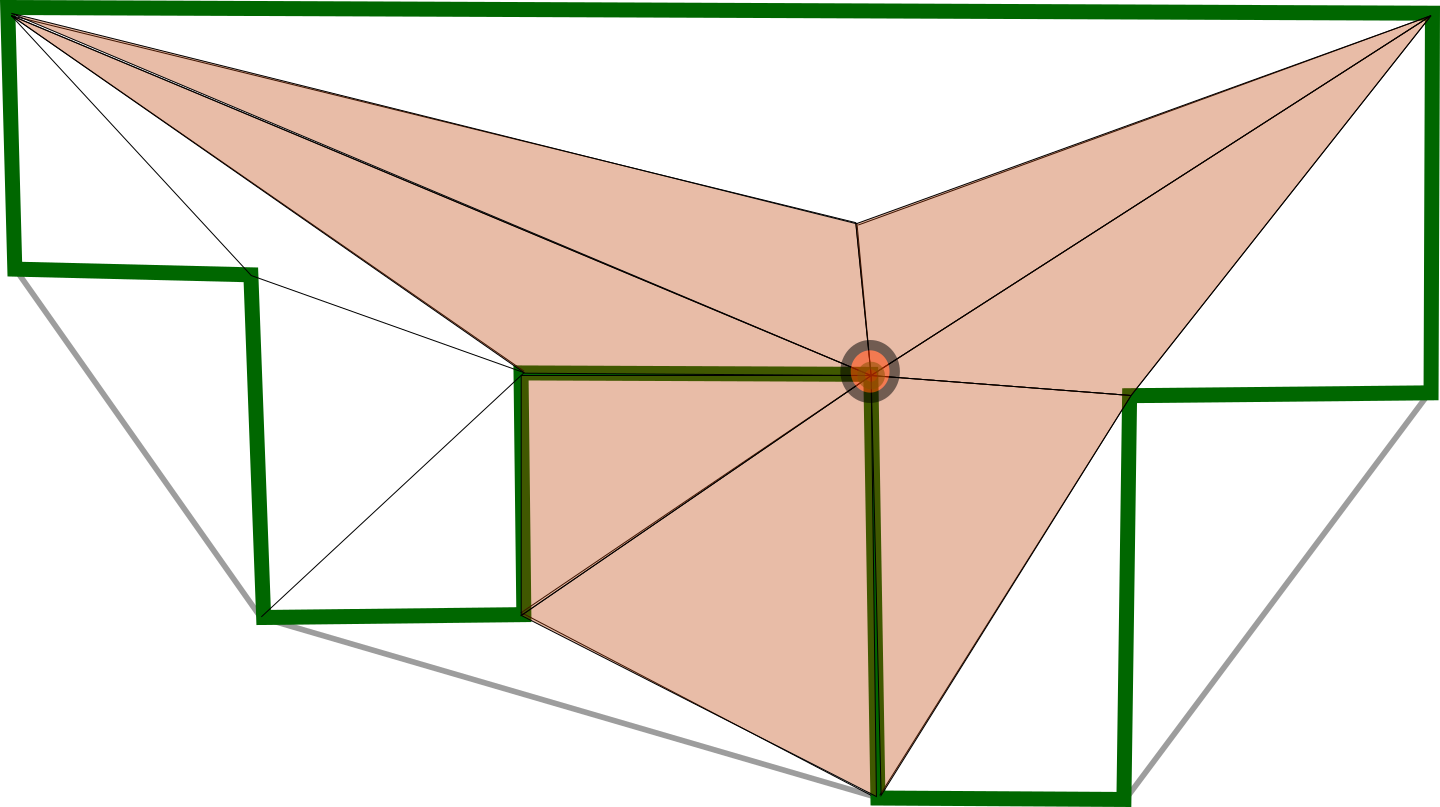}} \quad
	\subfigure[\label{fig:proc3} The disk at the first iteration.]
	{\includegraphics[width=4cm]{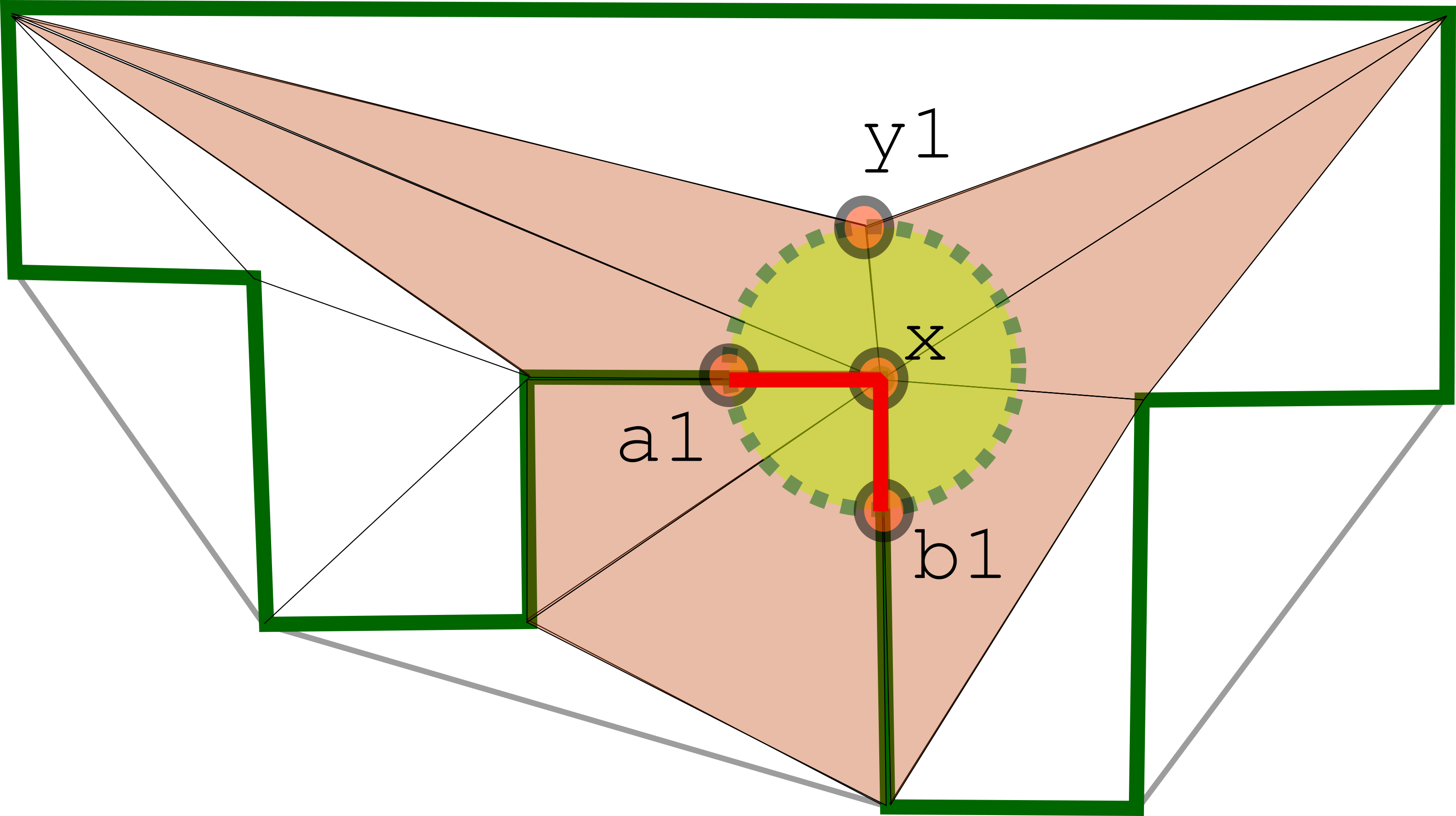}} \\
	\subfigure[\label{fig:proc4} $R(x)$ and the disk at the second iteration]
	{\includegraphics[width=4cm]{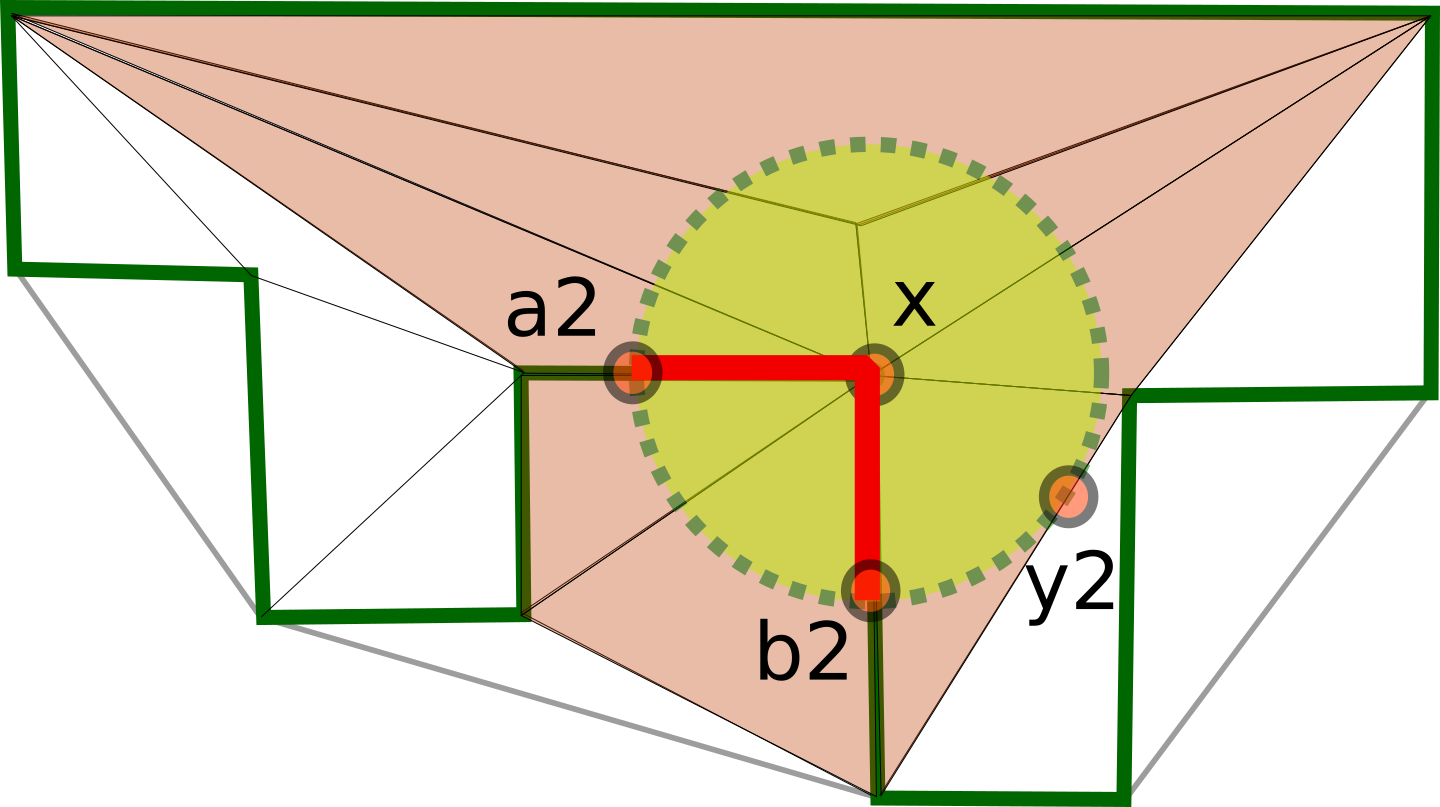}} \quad
	\subfigure[\label{fig:proc5} $R(x)$ and the disk at the third (and last) iteration]
	{\includegraphics[width=4cm]{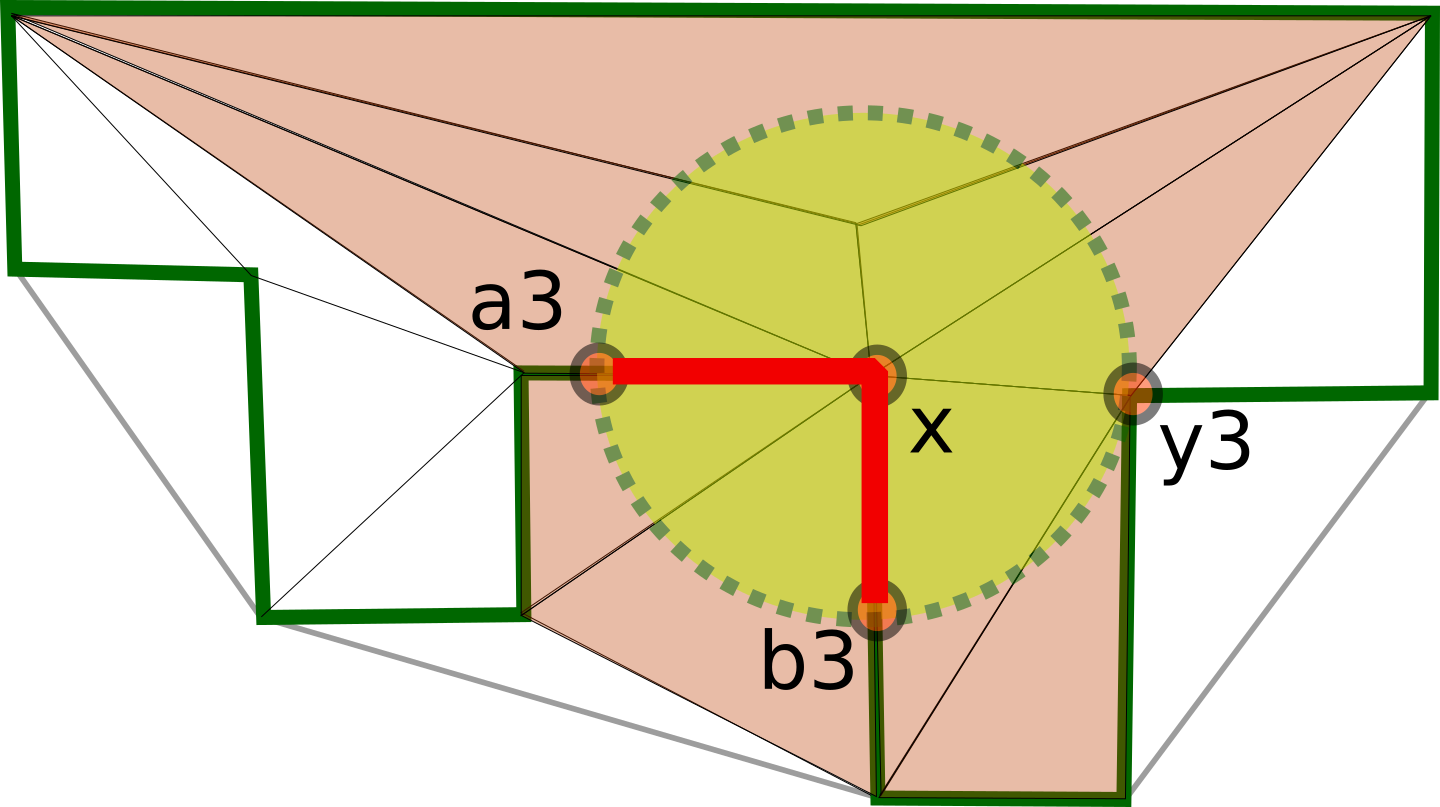}} \quad
	
	\caption{Computing $E$ for a single vertex. \subref{fig:proc1} Input polygon. Both the convex hull and the triangulation are shown. \subref{fig:proc2} At the first iteration, $R(x)$ includes all the triangles incident at $x$. \subref{fig:proc3} From the boundary of $R(x)$, the point $y_1$ closest to $x$ is selected. The subset of polygon elements within the disk is a single connected component (i.e. the open polyline $\overline{a_1xb_1})$. \subref{fig:proc4} $R(x)$ is updated by adding $y_1$'s top-most incident triangle, and the closest point $y_2$ is computed. Again, the subpart of the polygon within the disk is a single open polyline. \subref{fig:proc5} $R(x)$ grows, and the closest point $y_3$ is computed. At this iteration, the disk includes two disconnected components (i.e. the open polyline $\overline{a_3xb_3}$ and the point $y_3$). Thus, a topology change occurred, $E(x)$ is computed as the distance from $x$ to $y_3$, and the algorithm terminates. }
	\label{fig:v_procedure}
\end{figure*} 


In the remainder, $M$ denotes the constrained triangulaton of $P$, a \emph{mesh edge} is an edge of $M$, whereas a \emph{polygon edge} is an edge of $P$. So, a polygon edge is also a mesh edge but the opposite may not be true. Analogously, we refer to \emph{mesh vertices} and \emph{polygon vertices}, which is useful because the triangulation may (or may not) have additional vertices that do not belong to the original $P$. A generic \emph{polygon element} indicates either a polygon edge or a polygon vertex.

An example of this algorithm is shown in Figure \ref{fig:v_procedure}. Let $x$ be a vertex of $P$. Let $R(x)$ be our growing region which is initially empty, and $\partial R(x)$ be the boundary of this region.
Also, let $A(x) = e_1 \cup e_2$ bet the portion of $P$ made of the two polygon edges incident at $x$.
In the first iteration, $R(x)$ includes all the triangles incident at
$x$ (Figure \ref{fig:proc2}). Iteratively, the algorithm computes the point $y$ on $\partial
R(x) - A(x)$ which is closest to $x$. At any iteration there might be multiple
points at the same distance from $x$. In this case the algorithm
randomly selects one of them, and all the others are detected in the
subsequent iterations. Depending on the nature of the selected point $y$,
the algorithm may either grow the region or terminate by associating
$E(x)=d(x,y)$ to $x$, where $d(x,y)$ is the Euclidian distance between $x$ and $y$.

Specifically, the closest point $y$ may be a vertex of $\partial R(x)$ (Figures \ref{fig:proc3} and Figure \ref{fig:proc5}),
or it may be in the middle of one of its edges (Figure \ref{fig:proc4}). Notice that $y$ is not necessarily 
a polygon element. Imagine to build a disk centered in $x$ and whose
radius equals the distance from $x$ to $y$. The algorithm terminates if
$y$ is what we call the \emph{antipodean} of $x$, that is, $y$ is on
$\partial P$ and the topology of $\partial P$ restricted to the disk is
no longer a single open line due to the inclusion of $y$.
For this to happen, one of the following conditions must hold:
\begin{itemize}
	\item $y$ is a point in the middle of a polygon edge;
	\item $y$ is a polygon vertex and its two incident polygon edges are either both \emph{internal} or both \emph{external} wrt the disk. Herewith, one such edge is \textit{external} if $y$ is its closest point to $x$, whereas it is internal in the other cases.
\end{itemize}

If none of these conditions holds, the algorithm grows $R(x)$ by adding
all the triangles whose boundary contains $y$. Thus, if $y$ is a vertex,
$R(x)$ grows on all the triangles which are incident at $y$ but are not
in $R(x)$ yet. If $y$ is in the middle of an edge $e$, $R(x)$ includes
the triangle incident at $e$ which is not in $R(x)$ yet.


\subsection{Minimum Thickness for a Single Edge}
\label{sec:2d_edge_method} 
The region growing approach described in Section \ref{sec:2d_vertex_method} can be extended to compute the minimum $E$ along a single edge $e = \langle v_1, v_2 \rangle$. In this case, the minimum thickness may be either at one of the two endpoints or inbetween. Intuitively, an arbitrarily small disk is centered on $e$ and is grown iteratively by exploiting $M$ to discretize the process. The center of the disk is not fixed, but is moved along $e$ at each iteration. At the last iteration, the disk radius represents the minimum value of $E$ along $e$, while the position of the disk center shows where the local minimum lies.

\begin{figure}[!h]
	\centering
	\subfigure[\label{fig:distance_vertex}]
	{\includegraphics[trim={3cm 3cm 0 11cm}, clip, height=3cm]{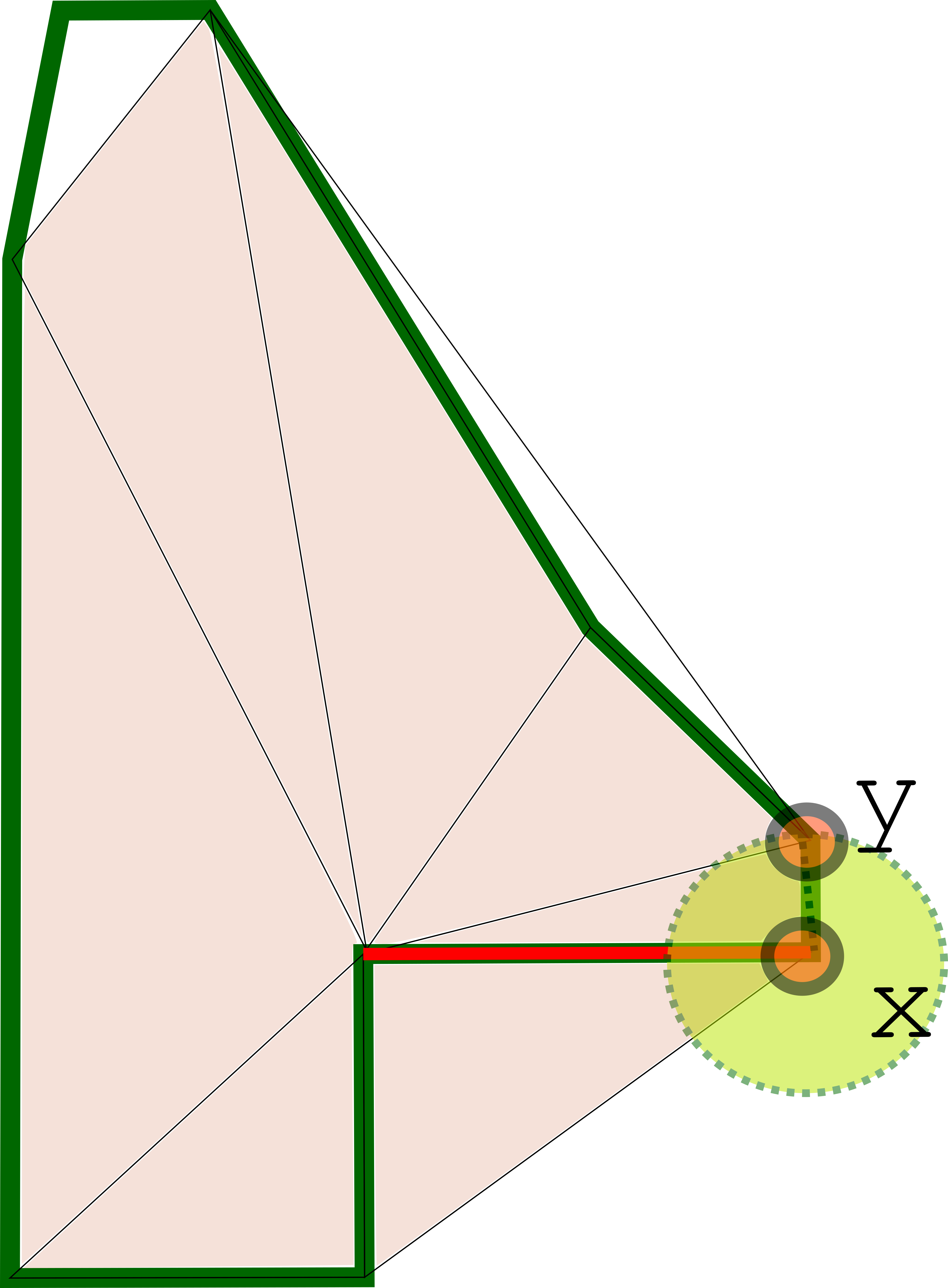}} \qquad
	\subfigure[\label{fig:distance_edge}]
	{\includegraphics[trim={3cm 3cm 0 11cm}, clip, height=3cm]{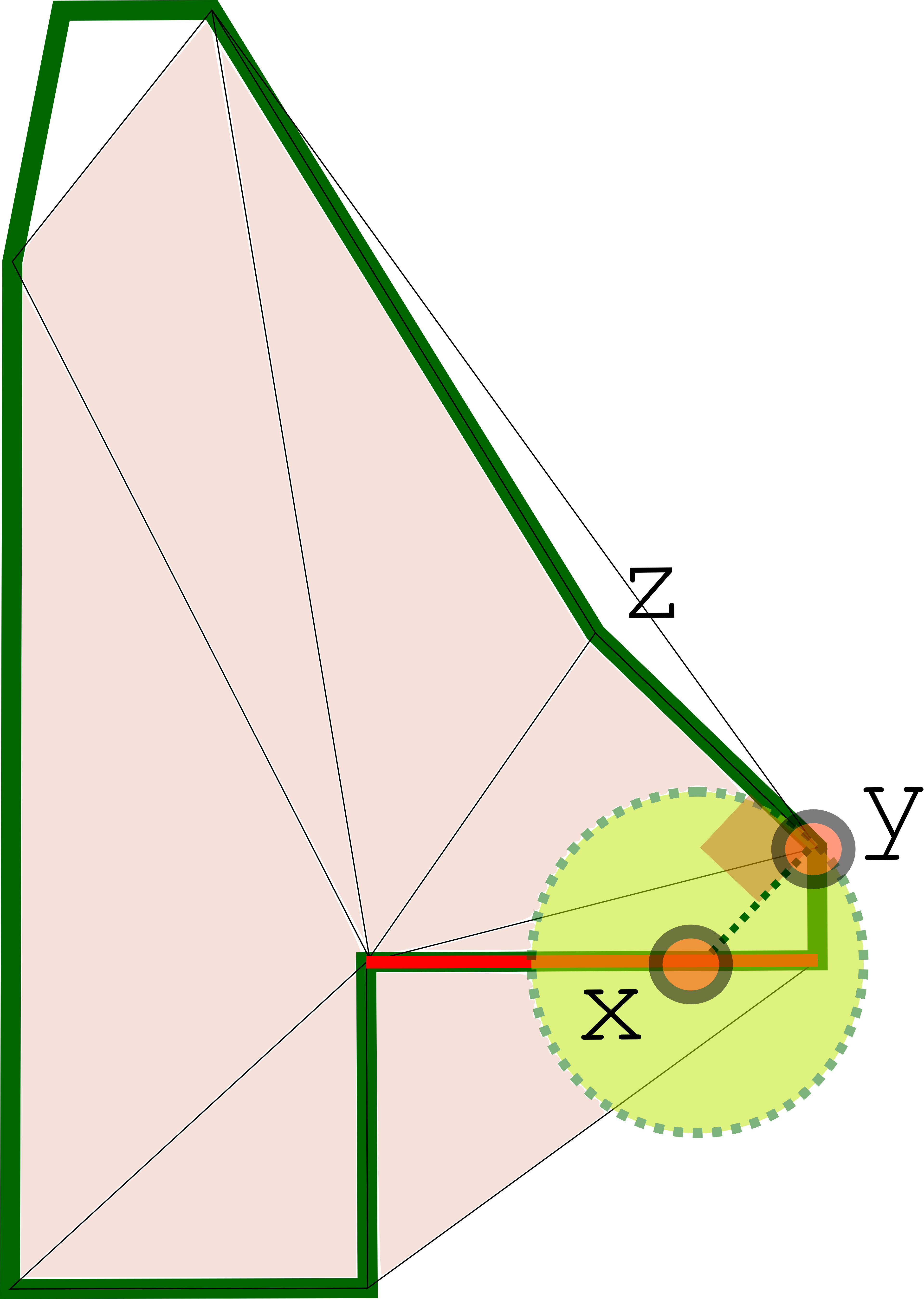}} \quad
	\caption{Two successive moments of the disk growth process while computing the minimum $E$ along an edge (red colored). \subref{fig:distance_vertex} The smallest disk centered on $e$ and tangent to vertex $y$. \subref{fig:distance_edge} The smallest disk centered on $e$ and tangent to edge $\overline{yz}$.}
	\label{fig:distances}
\end{figure}

\begin{figure*}[!h]
	\centering
	\subfigure[\label{fig:proc_edge1} Convex hull and triangulation.]
	{\includegraphics[width=3.2cm]{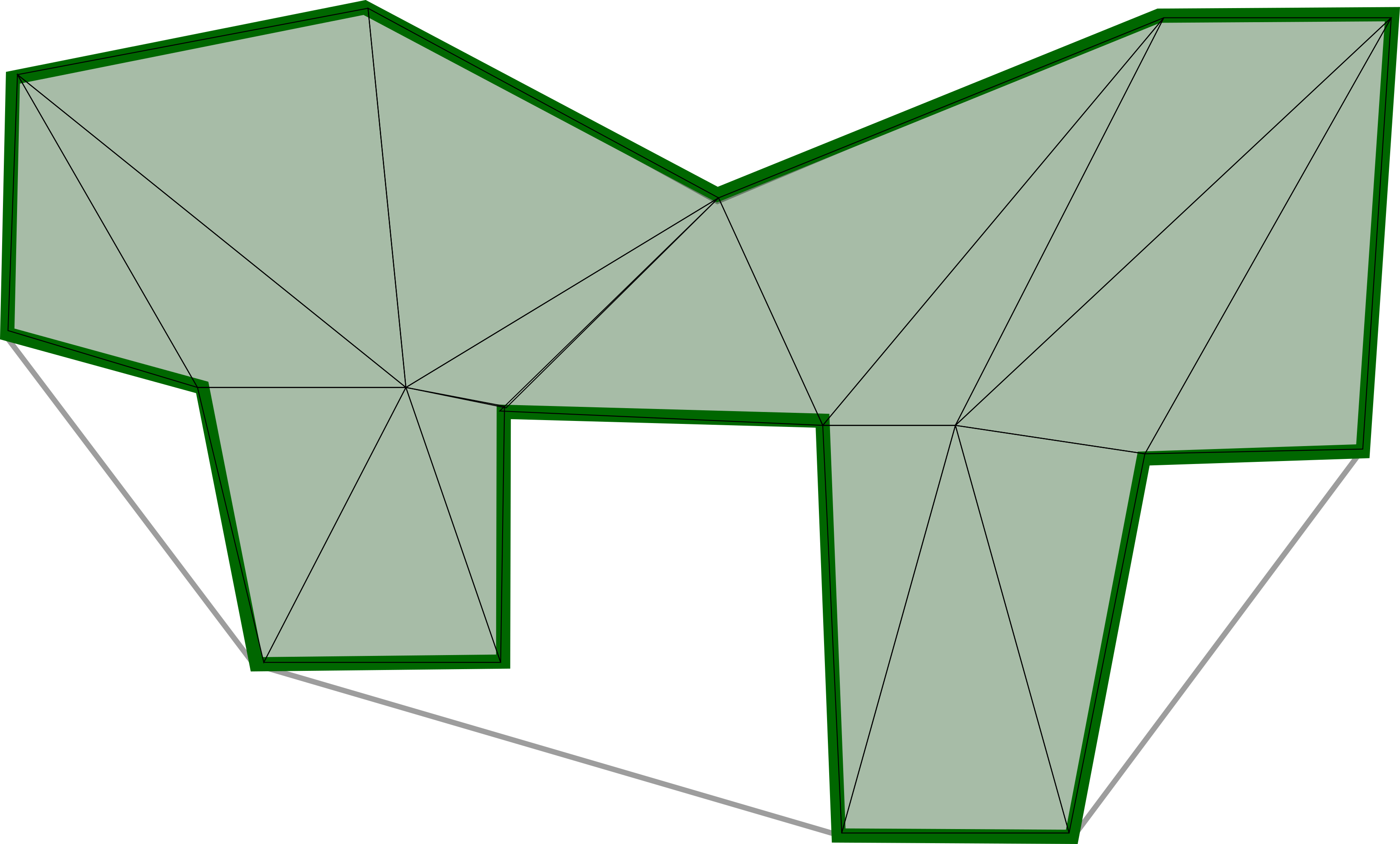}}\qquad
	\subfigure[\label{fig:proc_edge2} $R(e)$ at the first iteration.]
	{\includegraphics[width=3.2cm]{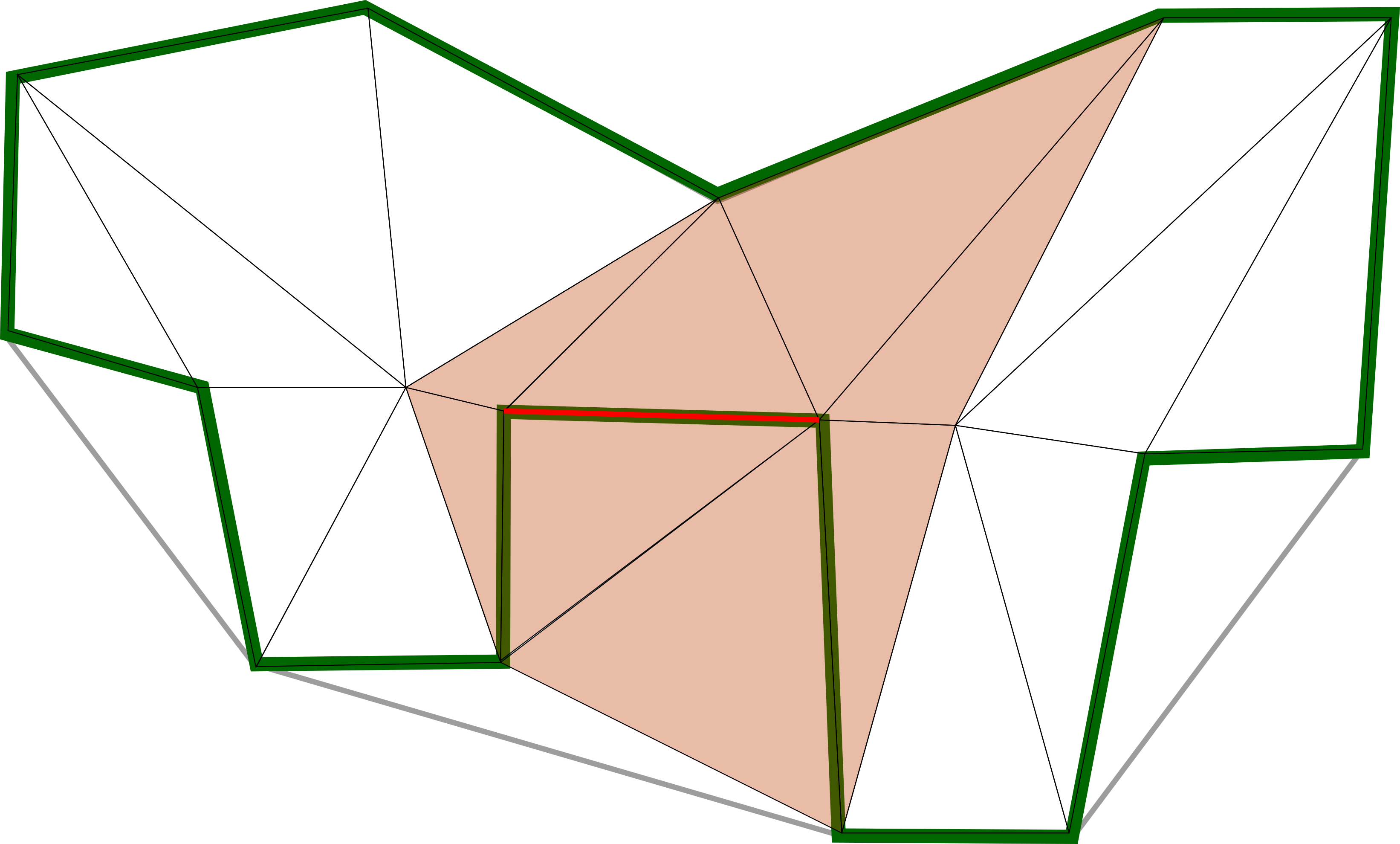}} \quad
	\subfigure[\label{fig:proc_edge3} The disk at the first iteration. ]
	{\includegraphics[width=3.2cm]{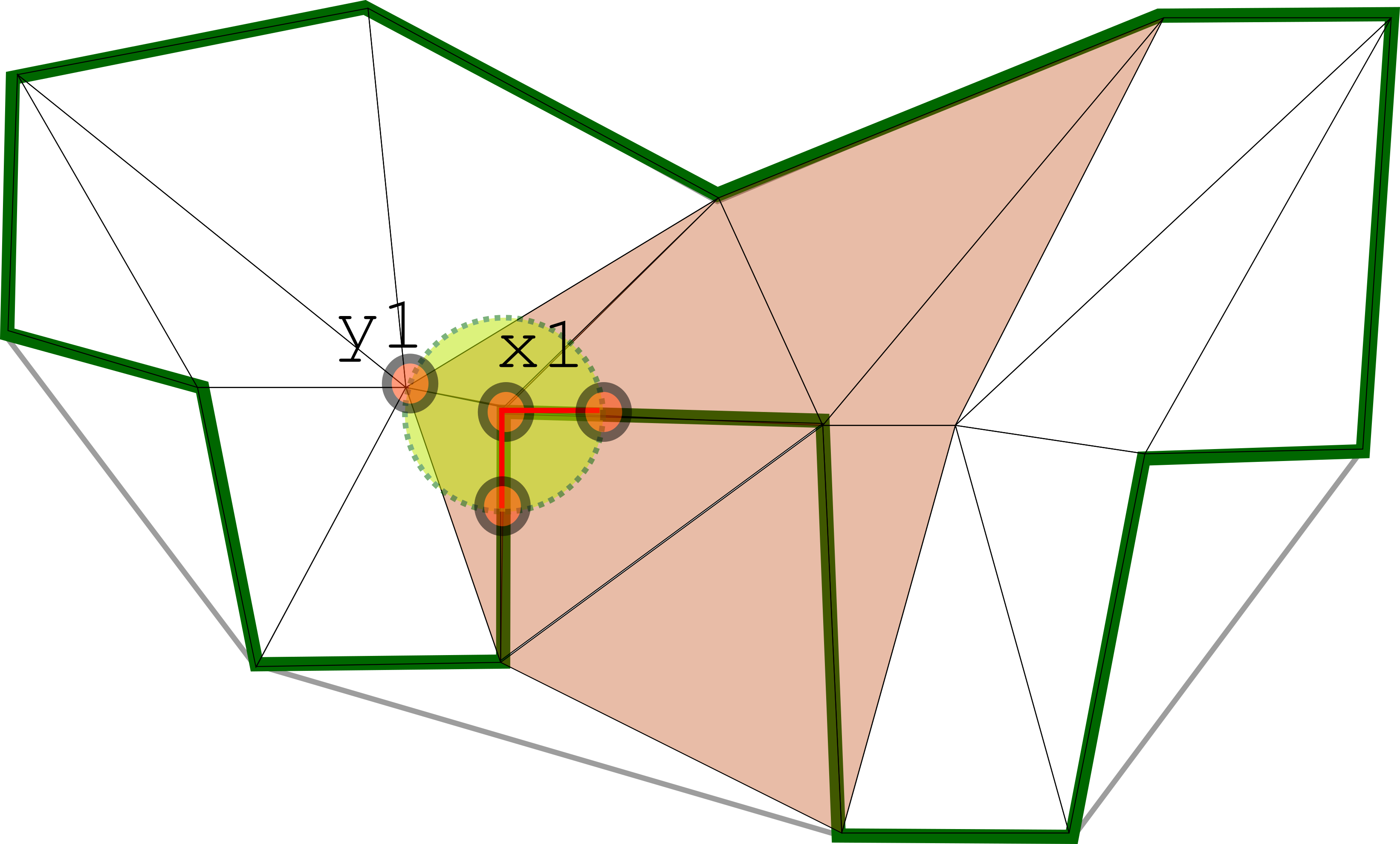}} \quad
	\subfigure[\label{fig:proc_edge4} $R(e)$ and the disk at the second iteration.]
	{\includegraphics[width=3.2cm]{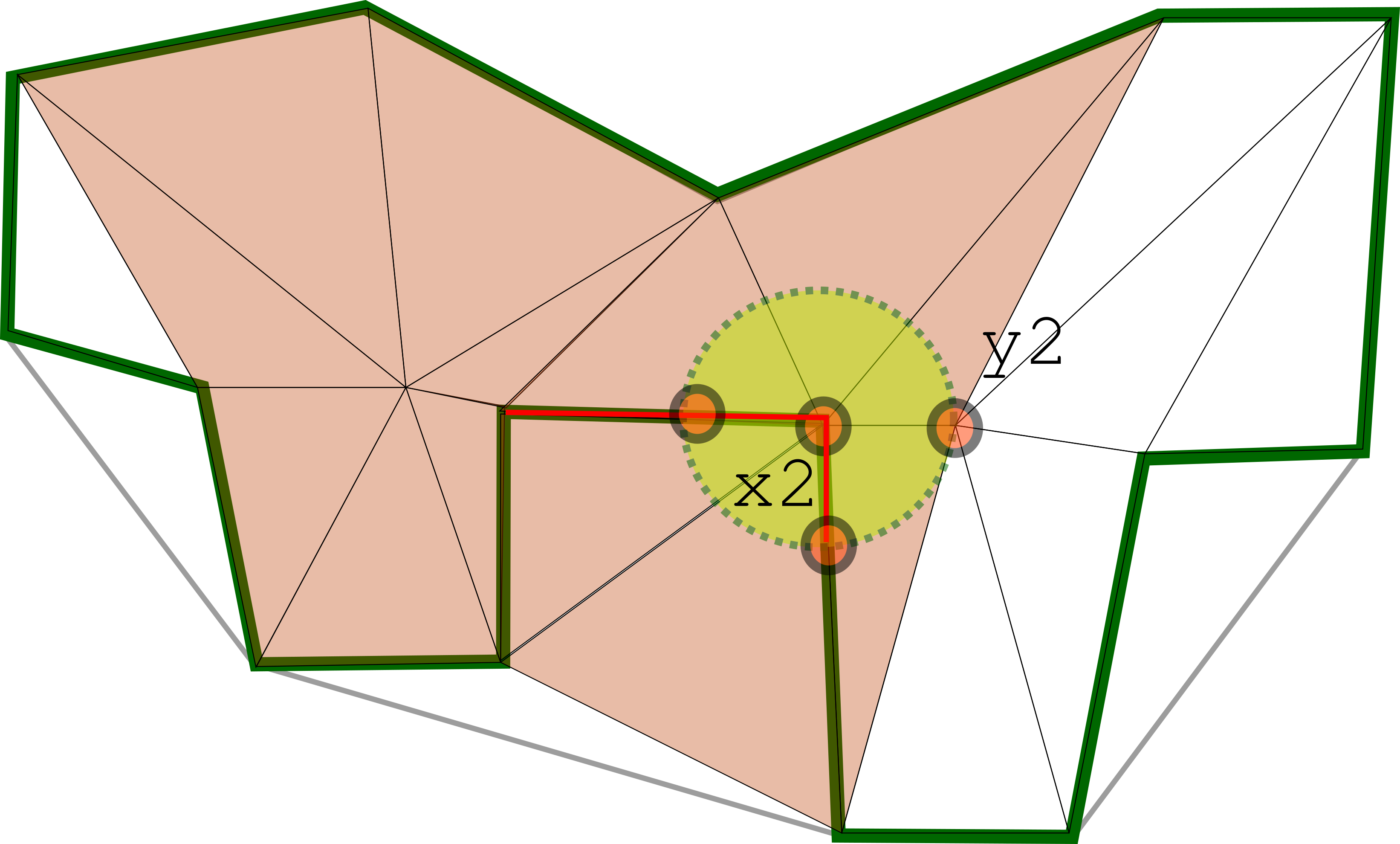}} \quad
	\subfigure[\label{fig:proc_edge5} $R(e)$ and the disk at the third iteration.]
	{\includegraphics[width=3.2cm]{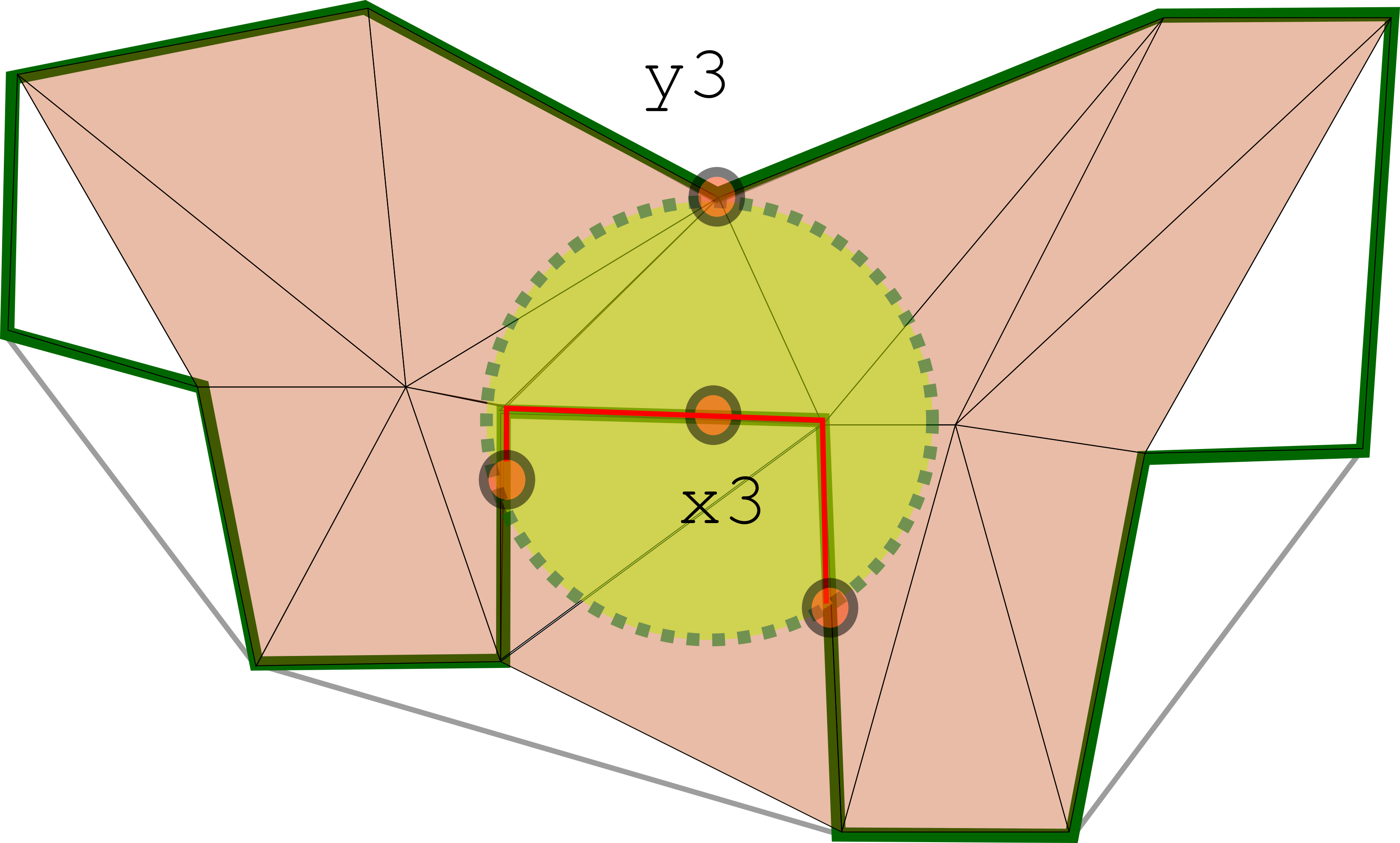}} 
	
	\caption{Computing the minimum $E$ along a single edge. \subref{fig:proc_edge1} The convex hull and the triangulation. \subref{fig:proc_edge2} The initial $R(e)$. \subref{fig:proc_edge3} Point $y_1$ is not the antipodean of $x_1$. $R(e)$ grows by including all the triangles incident at $y_1$. \subref{fig:proc_edge4} Second iteration. \subref{fig:proc_edge5} Third iteration.
	}
	\label{fig:procedure_edge}
\end{figure*}

Before starting the region growing, we perform an acute angle test: let $e_1$ be the polygon edge that shares $v_1$ with $e$. If the angle formed by $e$ and $e_1$ at $v_1$ is acute, we assign a zero value of $E$ to $v_1$. Similarly, we measure the angle at the other endpoint $v_2$ and set $E(v_2)$ to zero if such an angle is acute. If either of the endpoints have been set to zero, the algorithm terminates. Indeed, in this case the minimum for the edge has already been found.

Otherwise, the algorithm initializes a region $R(e)$ with all the triangles that share at least a vertex with $e$. Let $A(e)$ be the union of the two polygon edges different than $e$ that are incident at one of the endpoints of $e$. At each iteration, the point $y$ on $\partial R(e) - A(e)$ which is closest to $e$ is computed, and the corresponding point $x$ on $e$ which is closest to $y$ is determined. If $y$ is the ``antipodean'' of $x$ the algorithm terminates, otherwise it proceeds with the region growing as described in Section \ref{sec:2d_vertex_method}.
Notice that in this description we have implicitly extended the definition of antipodean of a vertex to any point on the boundary of $P$: now, $x$ is not necessarily a vertex, and this fact has a subtle though major impact on our process. Indeed, the algorithm should terminate when the disk becomes tangent to a polygon edge in the middle, or in any other case when $\partial P$ within the disk is no longer a single open line. But unfortunately, this is not sufficient. Consider the disk in Figure \ref{fig:distance_edge}: such a disk is tangent to the polygon edge $\overline{yz}$ in one of its endpoints and no termination condition holds. However, any arbitrarily small displacement of the disk along $e$ while possibly growing the disk itself would cause an event which is a stop condition of the algorithm (i.e. the disk would become tangent to $\overline{yz}$ in the middle). To turn this observation into practice and guarantee to detect these events, we say that a point $y$ on a polygon edge $e'$ of $\partial P$ is the antipodean of a point $x$ in the middle of an edge $e$ if the segment $x-y$ is orthogonal to $e'$.
Hence, our region growing algorithm terminates either if $y$ is a polygon edge endpoint that satisfies this orthogonality condition, or if it is a polygon vertex satisfying the conditions given in Section \ref{sec:2d_vertex_method}.

\subsection{Global thin feature detection}
\label{sec:2d_queue}
A comprehensive $\epsilon$-map can be built by splitting all the edges at local minima and by associating the value of $E(x)$ to each vertex $x$, based on the procedures described in Sections \ref{sec:2d_edge_method} and \ref{sec:2d_vertex_method}.

Specifically, if our procedure detects a local minimum in the middle of an edge, that edge is split, the triangulation is updated accordingly, and the algorithm is run again to compute the local minima along the two resulting sub-edges. This procedure is guaranteed to converge because $P$ is piecewise-linear and $E(P)$ must necessarily have a finite number of local minima. Particular/degenerate cases such as, e.g. parallel edges, are avoided through Simulation of Simplicity \cite{Edelsbrunner:1990}.

\begin{figure}[!ht]
	\centering
	\includegraphics[width=0.3\textwidth]{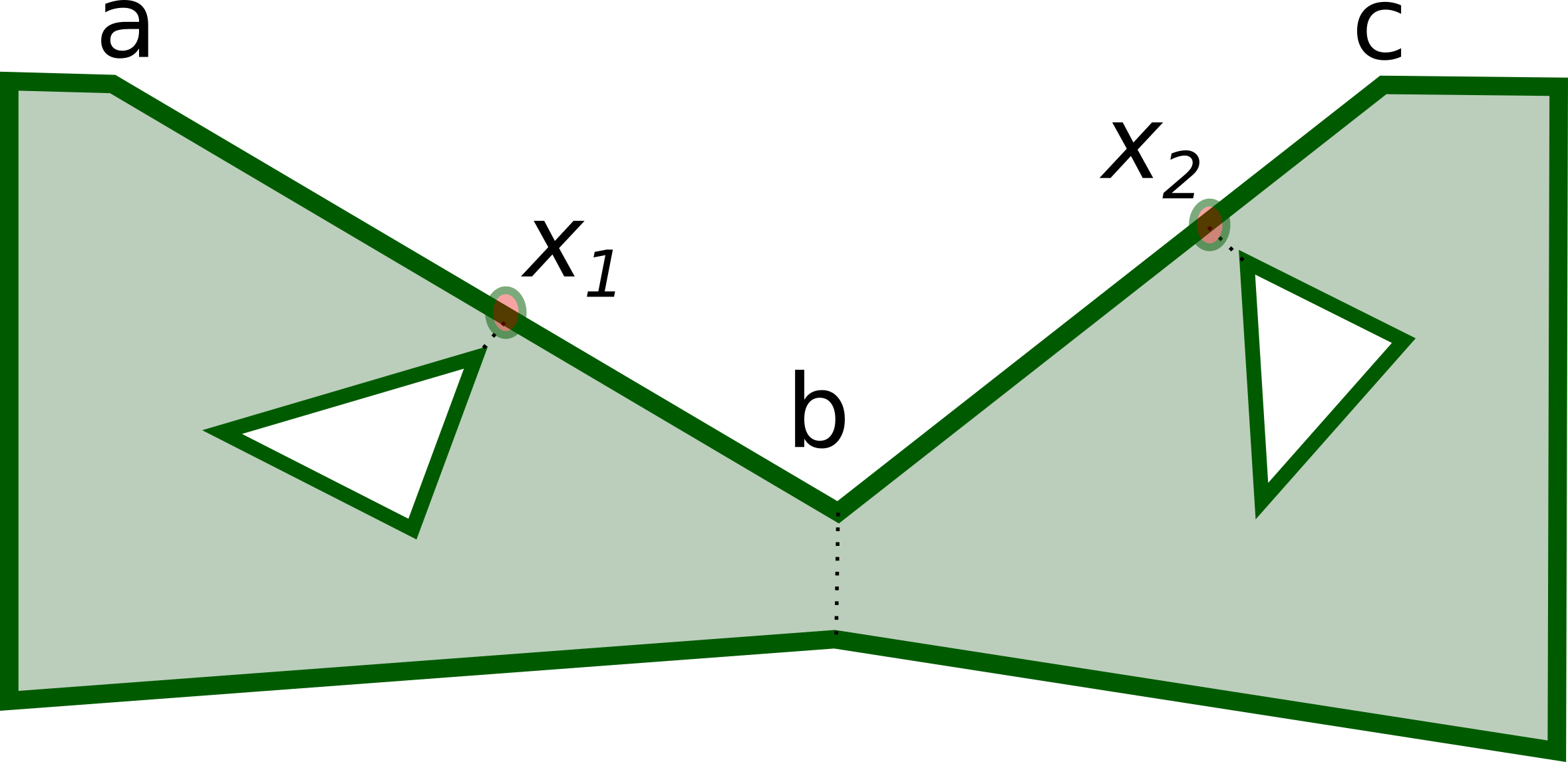}
	\caption{Computing the minima $E$ along any polygon edge is not sufficient. In this example, $x_1$ and $x_2$ are the minima $E$ along the edge $\overline{ab}$ and $\overline{bc}$ respectively. Recursively, the generated sub-edges $\overline{ax_1}$, $\overline{x_1b}$, $\overline{bx_2}$ and $\overline{x_2c}$ are analyzed. $x_1$ is the minimum $E$ along the first two sub-edges. $x_2$ is the minimum $E$ along the other two sub-edges. The value of $E$ at vertex $b$ is unknown, but it might be less than the given threshold. Thus, it is necessary to compute $E(b)$.}
	\label{fig:example1}
\end{figure}

One might argue that computing $E$ at edges only is sufficient because any vertex is the end-point of some edges, and therefore its value of $E$ would be computed by the procedure in Section \ref{sec:2d_edge_method}.
Unfortunately this is not true for all the vertices, and a counter-example is shown in Figure \ref{fig:example1}. Thus, the algorithm described in Section \ref{sec:2d_vertex_method} must be necessarily run to compute $E$ at any polygon vertex which is not a local minimum for any edge. Only after this step the $\epsilon$-map is guaranteed to be correct at all the vertices.

We observe that many applications need to detect and possibly process thin features only, while thickness information on other (thick enough) parts of the shape is not used (Section \ref{sec:thickening}). 
In these cases our algorithm can be significantly accelerated by specifying a threshold thickness beyond which a feature is no longer considered to be thin, and by stopping the region growing as soon as the radius exceeds such a threshold value.
Hence, in this case the resulting $\epsilon$-map exactly represents the thickness at any relevant point bounding a thin feature, whereas a generic \emph{thick} attribute may replace the value of $E$ for all the other boundary points of the input shape.

\section{Solid Object Analysis}
\label{sec:3d}

In 3D we analyze a polyhedron $P$ bounded by a 2-manifold triangle mesh $M$, and use a
constrained tetrahedrization $T$ of $P$'s convex hull to discretize the
ball-growing process.

We observe that the minimum of the $\epsilon$-map within a triangle can be either on its border or in the middle of its internal area. Thus, before computing $E$ at vertices and possibly splitting edges as we do in 2D, in 3D we may need to split triangles as well to represent some of the local minima.

Vertices, edges and triangles are qualified as \emph{polyhedral} if they
belong to $M$, whereas they are \emph{mesh elements} if they belong to $T$.
Thus, similarly to the 2D case, a polyhedral element is also a mesh
element, but not necessarily vice-versa.

\subsection{Thickness at a vertex}
\label{sec:3d_method_vertex}
We proceed with the region growing as we do for the 2D case,
with the difference that in this case $R(x)$ is a tetrahedral mesh and
its boundary is a triangle mesh. Thus, $A(x)$ is the union of all the polyhedral triangles incident at $x$,
$R(x)$ is initialized with the 
set of all the tetrahedra incident at $x$, and
the algorithm terminates if the closest point $y$ on $\partial R(x) - A(x)$ is
the antipodean of $x$, that is, when the topology on $\partial P$
restricted to the ball is no longer a single disk.
To define the conditions that characterize an antipodean point, we qualify
a point $y$ on $\partial R(x)$ based on its incident elements as follows.

If $t$ is a polyhedral triangle having $y$ on its boundary (either along 
an edge or on one of its three vertices), $t$ may be either completely out of the ball
(i.e. its minimum distance from $x$ is exactly the ball's radius), or it
may be partly or completely contained in it (i.e. its minimum distance
from $x$ is smaller than the ball's radius). In the former case we say
that $t$ is \emph{external}, whereas in the latter case we say that $t$ is an
\emph{internal} triangle wrt the ball. We use an analogous terminology to
characterize an edge incident at $y$ wrt to the ball.

Having said that, $y$ is the antipodean of $x$ if one of the following
conditions holds:
\begin{itemize}
	\item $y$ is in the middle of a polyhedral triangle;
	\item $y$ is in the middle of a polyhedral edge whose two incident polyhedral
	triangles are either both external or both internal;
	\item $y$ is a polyhedral vertex whose incident polyhedral edges are $e_1, ...,
	e_n$ in radial order around $y$ and either:
	\subitem all the $e_i$'s are internal or
	\subitem all the $e_i$'s are external or
	\subitem there are more than two switches in the ordered chain of the
	$e_i$'s (internal/external or vice-versa, including the possible switch
	between $e_n$ and $e_1$).
\end{itemize}

Thus, at the end of each iteration, we check whether $y$ satisfies one of the aforementioned conditions.
If so, the algorithm terminates.
Otherwise, we grow the region on all the tetrahedra having $y$ on the boundary and not being already part of R(x).

\subsection{Minimum thickness along an edge}
\label{sec:3d_method_edge}
During the analysis of a polyhedral edge $e = \langle v_1, v_2 \rangle$ we consider the possibility to move the ball's center along the edge. The position of the ball's center at the last iteration represents the minimum of $E$ along the edge, while the radius indicates its actual value. Note that the minimum may correspond to an edge endpoint or may lie in the middle.

Before starting the iterative region growing, we check for acute configurations as follows. Let $t_1$ and $t_2$ be the two polyhedral triangles incident at $e$. We consider all the edges which are incident at $v_1$ but are not edges of either $t_1$ or $t_2$. If any such edge forms an acute angle with $e$, $E(v_1)$ is set to zero, otherwise we consider all the triangles incident at $v_1$ but not incident at $e$. If the projection of $e$ on the plane of any such triangle intersects the interior of the triangle itself, $E(v_1)$ is set to zero. Analogous checks are performed on $v_2$. If either $E(v_1)$ or $E(v_2)$ is set to zero due to these checks, the algorithm terminates.

Otherwise, a region $R(e)$ is initialized with all the tetrahedra that share at least a vertex with $e$, and the set $A(e)$ is made of all the polyhedral triangles that share at least a vertex with $e$.
At each iteration, we calculate the point $y$ on $\partial R(e) - A(e)$ which is closest to $e$, and determine the corresponding point $x$ on $e$ which is closest to $y$.
If $y$ is the "antipodean" of $x$ the algorithm terminates, otherwise it proceeds with the region growing as described in Section \ref{sec:2d_vertex_method}.

Similarly to the 2D case, the definition of antipodean was implicitly extended and a clarification is necessary to avoid missing borderline cases. Imagine a ball centered on $e$ and tangent to a polyhedral triangle. The tangency point may be either in the interior of the triangle or on its boundary. In the former case the algorithm terminates, whereas in the latter the region growing would take place if we consider only the conditions given in Section \ref{sec:3d_method_vertex}. Nevertheless, if we apply an arbitrarily small translation of the ball center along $e$ while possibly growing its radius, the ball would become tangent to the same triangle in the middle (i.e. the algorithm should terminate). Once again, to cope with these cases we extend the definition of antipodean point, and we say that a point $y$ on a polyhedral element $y'$ of $\partial P$ is the antipodean of a point $x$ in the middle of an edge $e$ if the segment $x-y$ is orthogonal to $y'$. Note that $y'$ may be either an edge or a triangle.

Hence, our region growing algorithm terminates if $y$ is on a polyhedral element that satisfies this orthogonality condition, or if it is a polyhedral vertex satisfying the conditions given in Section \ref{sec:2d_vertex_method}. 

\subsection{Minimum thickness on a triangle}
\label{sec:3d_method_triangle}
A similar approach can be exploited to compute the minimum $E$ on a single polyhedral triangle $t = \langle v_1, v_2, v_3 \rangle$. The minimum may correspond to a triangle vertex, or lie in the middle of a triangle edge, or be a point in the middle of the triangle itself. Thus, the ball's growth process considers the possibility to move the center on the whole $t$. 

As we do for edges, before starting the iterative region growing we check for acute configurations as follows. Let $t_1$, $t_2$ and $t_3$ be the three polyhedral triangles adjacent to $t$. We consider all the polyhedral edges which are incident at $v_1$ but are not edges of either $t_1$, $t_2$ or $t_3$. If the projection of any such edge on the plane of $t$ intersects the interior of $t$, $E(v_1)$ is set to zero. Analogous checks are performed on $v_2$ and $v_3$. 
Furthermore, we consider each polyhedral triangle $t_i$ that shares at least a vertex with $t$, and if the angle formed by the normal at $t$ and the normal at $t_i$ is obtuse, the value of $E$ for all their shared vertices is set to zero.
If the value of $E$ for at least one of the three vertices is set to zero due to these checks, the algorithm terminates. 

Otherwise, the initial region $R(t)$ includes all the tetrahedra which are incident to at least one vertex of $t$, and the set $A(t)$ is made of all the polyhedral triangles that share at least a vertex with $t$. 
At each iteration, we calculate the point $y$ on $\partial R(t) - A(t)$ which is closest to $t$, and determine the corresponding point $x$ on $t$ which is closest to $y$.
The same arguments and procedure described in Section \ref{sec:3d_method_edge} apply here.

\subsection{Global thin feature detection}
The algorithm to compute the overall $\epsilon$-map is similar to the 2D version described in Section \ref{sec:2d_queue}.
We first analyze each polyhedral triangle (Section \ref{sec:3d_method_triangle}), possibly split it at its local minimum, and re-iterate the analysis on the resulting sub-triangles. We observe that these sub-triangles can be either three or just two (the latter case occurs if the minimum is on one of the three edges).
When all the triangles are processed, we analyze each polyhedral edge (Section \ref{sec:3d_method_edge}), possibly split it at its local minimum, and re-iterate on the two resulting sub-edges.
The tetrahedization is updated upon each split operation performed on $P$.
Finally, we process all the polyhedral vertices (Section \ref{sec:3d_method_vertex}).

Note that in the 2D case we might have assumed that all the vertices are polygonal vertices. Indeed, a planar polygon can always be triangulated.
A corresponding statement cannot be done for the 3D case due to the existence of polyhedra that do not admit a tetrahedrization. In these cases, a constrained tetrahedrization can be calculated only if a number of so-called Steiner points are added to the set of input vertices.

\section{Thickening}
\label{sec:thickening}

When thickness information is available, the overall geometry can be adaptively changed in an attempt to replace thin features with thicker structures. For some applications a minimum thickness may be well defined (e.g. the layer thickness for 3D printing), whereas for some others this value can be empirically set by the user based on his/her experience.
In any case, in this section we assume that such a threshold value is available and exploit it, in combination with thickness information, to modify the shape locally. While doing this operation, we strive to modify the geometry only where necessary, and no more than necessary.

Our thickening problem can be formulated as follows: given a shape $S$ and a threshold thickness $\epsilon$, we wish to find the shape $S'$ which is most similar to $S$ while being an $\epsilon$-shape. Clearly, if $S$ has no feature thinner than $\epsilon$, then $S'$ must coincide with $S$.
Though an exact solution to this problem appears to be extremely complicated, herewith we present a heuristic approach that proved to be both efficient and effective in all of our experiments. The basic idea is the following: if the value $E(x)$ of the $\epsilon$-map at a point $x$ on $S$ is less than the threshold $\epsilon$, we consider both $x$ and its antipodean $y$, and move both the points away from each other as long as their distance becomes equal to $\epsilon$. Stated differently, the new position for $x$ will be $x + \frac{(\epsilon-E(x))}{2} \overline{x-y}$, whereas the new position for $y$ will be $y + \frac{(\epsilon-E(x))}{2} \overline{y-x}$,
where $\overline{a}$ denotes the normalized vector $\frac{a}{||a||}$.

Since the objective in this section is to thicken too thin features, herewith we consider the positive $\epsilon$-map $E^{+}$ only. On piecewise-linear shapes, this map can be computed as described in Sections \ref{sec:2d} and \ref{sec:3d} by disregarding the outer part of the constrained triangulation/tetrahedrization, and by considering internal angles only when checking for acuteness.
Furthermore, a partial $\epsilon$-map can be computed as described in Section \ref{sec:2d_queue}. Indeed, by using $\epsilon$ as a threshold to stop the process, we can achieve a much faster computation.

After such a computation, each vertex can be categorized in three ways, depending on the value of its partial $\epsilon$-map: in the remainder, an \emph{acute vertex} is a vertex mapped to a zero value, while a \emph{thin vertex} is a vertex mapped to a positive value which is lower than the thickness threshold $\epsilon$. Any other vertex is just \emph{thick} and is not considered by the algorithm.
While computing the $\epsilon$-map we keep track of the antipodean point for each thin vertex. If such a point does not coincide with a vertex, before proceeding with the thickening we split the simplex that contains it, so as to have a vertex to displace.

For the sake of simplicity, we now assume that there are no acute vertices. Their treatment is described later in Sections \ref{sec:angle_2d} and \ref{sec:angle_3d}. Thus, each thin vertex $x$ with antipodean $y$ is associated to a displacement vector $\delta x = \frac{(\epsilon-E(x))}{2} \overline{x-y}$. Similarly, each antipodean vertex $y$ of $x$ is associated to a displacement vector $\delta y = \frac{(\epsilon-E(x))}{2} \overline{y-x}$. If the same vertex has several ``roles'' (e.g. it is the antipodean of two different vertices), its displacement vectors are summed.

When this displacement vector field is complete, the actual displacement may take place. However, even this operation must be undertaken with a particular care. Indeed, if two thin features are separated by an insufficient space, their uncontrolled growth might bring one to intersect the other. To avoid this, we keep the original triangulation/tetrahedrization (outer parts included) and, for each single displacement, we check that no flip occurs among the simplexes incident at the displaced vertex.
We observe that the need for such a check reveals that our problem may have no solution if we do not allow topological changes. However, if topological changes are acceptable, we just let the surfaces intersect with each other and track the so-called outer hull in a second phase as described in \cite{Attene:2014}.

The following two subsections describe how to pre-process the model so as to remove possible acute vertices and make the model ready to be thickened as proposed.

\subsection{Pre-processing 2D shapes}
\label{sec:angle_2d}
\begin{figure}[!h]
	\centering
	\includegraphics[width=0.45\textwidth]{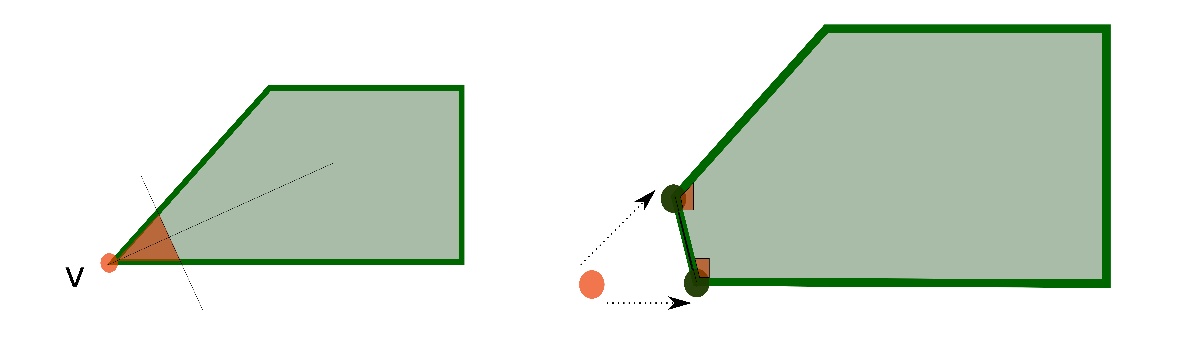} \qquad

	\caption{Removing acute angles. A cutting line is exploited which is perpendicular to the internal bisector of the angle.}
	\label{fig:cutting_line}
\end{figure}

Clearly, our thickening approach is not suitable for acute vertices which are mapped to a zero value and for which no antipodean point exists. Thus, preprocessing is performed to remove acute vertices before thickening. To remove acute angles, we imagine to cut off a small portion of the shape around each acute vertex, and to fill the generated hole. Specifically, let $v$ be an acute vertex. A cutting line is defined, which intersects the internal angle bisector and is perpendicular to it. The distance between $v$ and the cutting line may be arbitrarily small. Then, for each edge $e_i = \langle v , v_i \rangle $ incident at $v$,  $v$ is replaced by the intersection point between the cutting line and $e_i$. Finally, an additional edge is added to close the polygon (Figure \ref{fig:cutting_line}). Then, the two algorithms described in Section \ref{sec:2d_vertex_method} and \ref{sec:2d_edge_method} are exploited to update the $\epsilon$-map with thickness information related to the modified parts of the shape.

\subsection{Pre-processing 3D models}
\label{sec:angle_3d}
\begin{figure}[!h]
	\centering
	\includegraphics[width=0.45\textwidth]{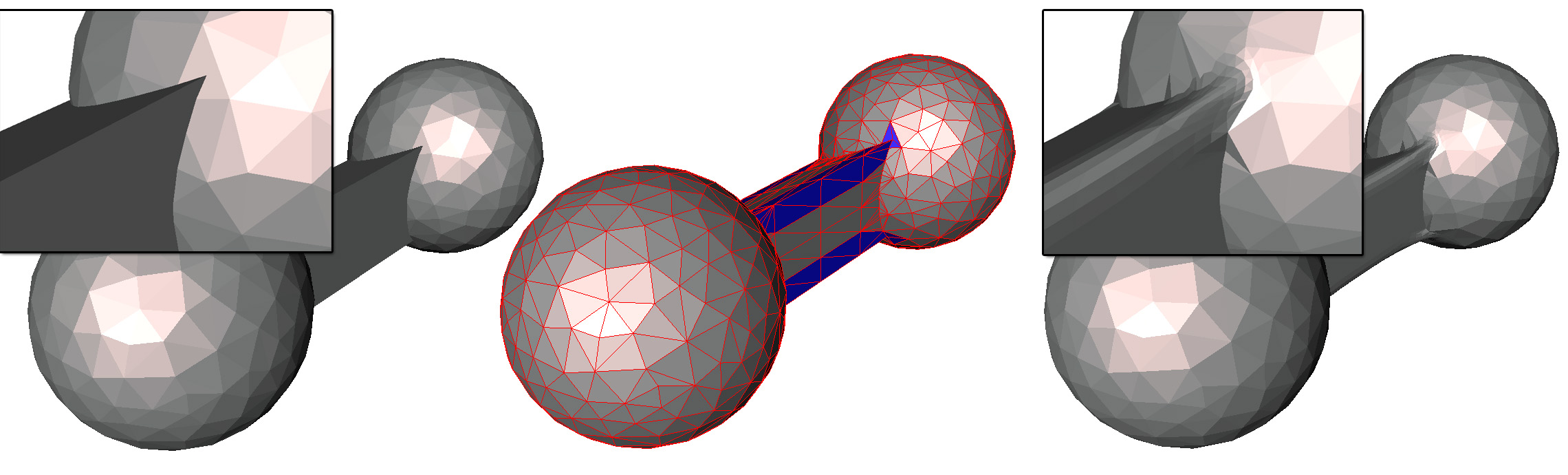}
	\caption{A model with acute vertices (left) and the triangles incident at these vertices (middle). After local subdivision, all the vertices are non-acute (right).}
	\label{fig:subdivision}
\end{figure}

Getting rid of acute vertices in 3D is not as easy as in 2D due to the possible presence of saddles. To solve this problem, we just consider the set of all the triangles which are incident at acute vertices and, on such a set only, we adaptively run a subdivision step using the modified interpolating Butterfly scheme by Zorin and colleagues \cite{Zorin:1996}. Then, the three methods proposed in Section \ref{sec:3d} are exploited to update the $\epsilon$-map restrictedly to the modified part of the model. The process is repeated as long as acute angles are found, and convergence is guaranteed because the subdivision generates a $C^1$ continuous surface at the limit (Fig. \ref{fig:subdivision}). To reduce the area of the modification, before running the aforementioned procedure we subdivide the considered triangles without changing the geometry (i.e. each edge is split at its midpoint) and keep working only on the subtriangles which are incident at acute vertices. This can be done as many times as necessary to keep the changes within a tolerable limit.

\section{Results and Discussion}
\label{sec:results}

We implemented both our analysis and thickening methods in C++ with the support of Tetgen \cite{Hang:2015} for the computation of the constrained tetrahedrizations. 
This section reports the results of some of our experiments on a set of 3D input meshes coming from \cite{DSW}.
Our prototype software provides the possibility to set an analysis direction, that is, the user can choose whether to compute a positive, a negative, or a bi-directional $\epsilon$-map.
Also, if the complete $\epsilon$-map is not necessary, the user can set a threshold value to compute only a partial $\epsilon$-map. In the latter case the region growing is interrupted as soon as the ball radius reaches the threshold.

\begin{figure}[!h]
	\centering
	
	\subfigure[\label{fig:map197} Model 197] {\includegraphics[width=0.13\textwidth]{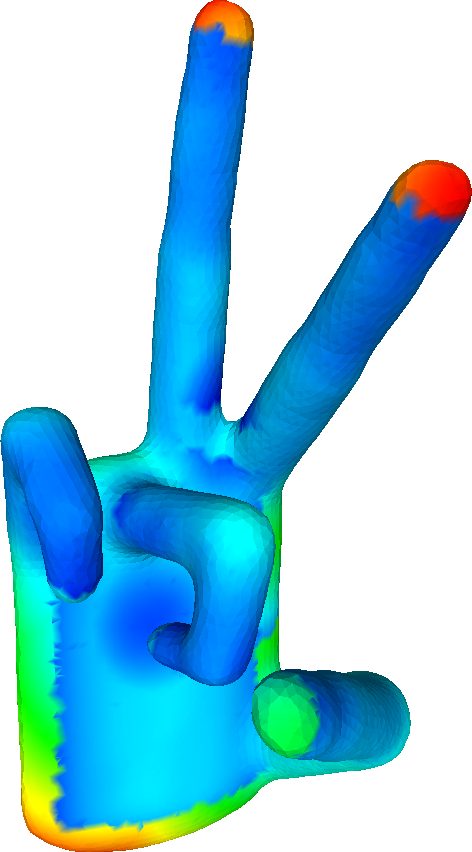}} \quad
	\subfigure[\label{fig:map121} Model 121] {\includegraphics[width=0.25\textwidth]{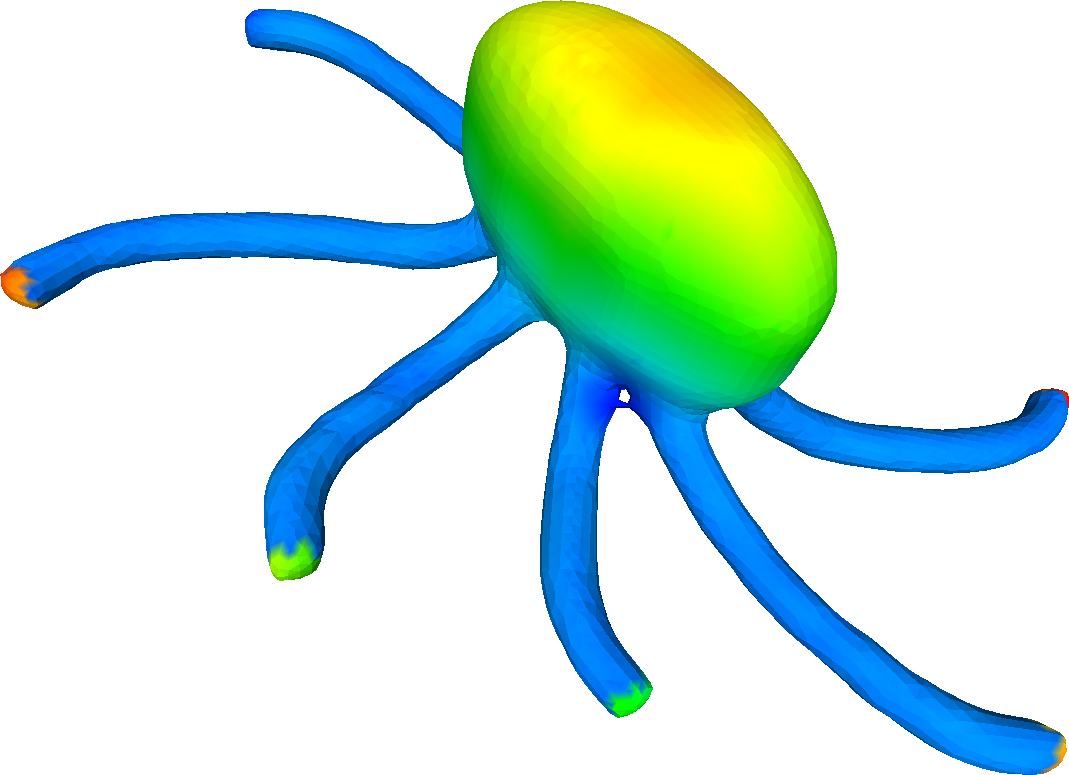}}  \quad
	
	\caption{
		Complete $\epsilon$-maps. The color palette goes from blue (thinnest parts) to red (thickest parts).
	}
	\label{fig:maps}
\end{figure}

\begin{figure}[!h]
	\centering
	
	
	\subfigure[\label{fig:positive_all}]
	{
		\includegraphics[width=0.1\textwidth, angle=90]{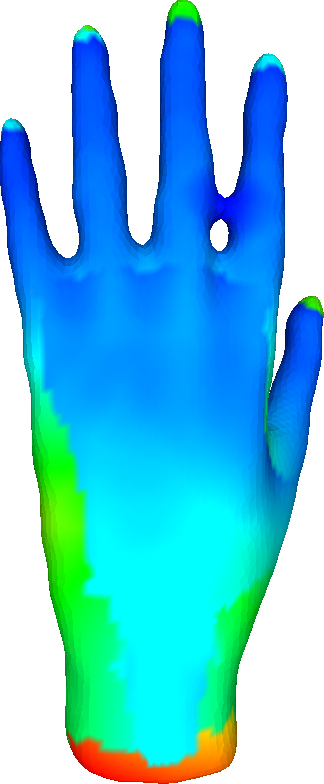}
	} \quad
	\subfigure[\label{fig:positive_thin}]
	{
		\includegraphics[width=0.1\textwidth]{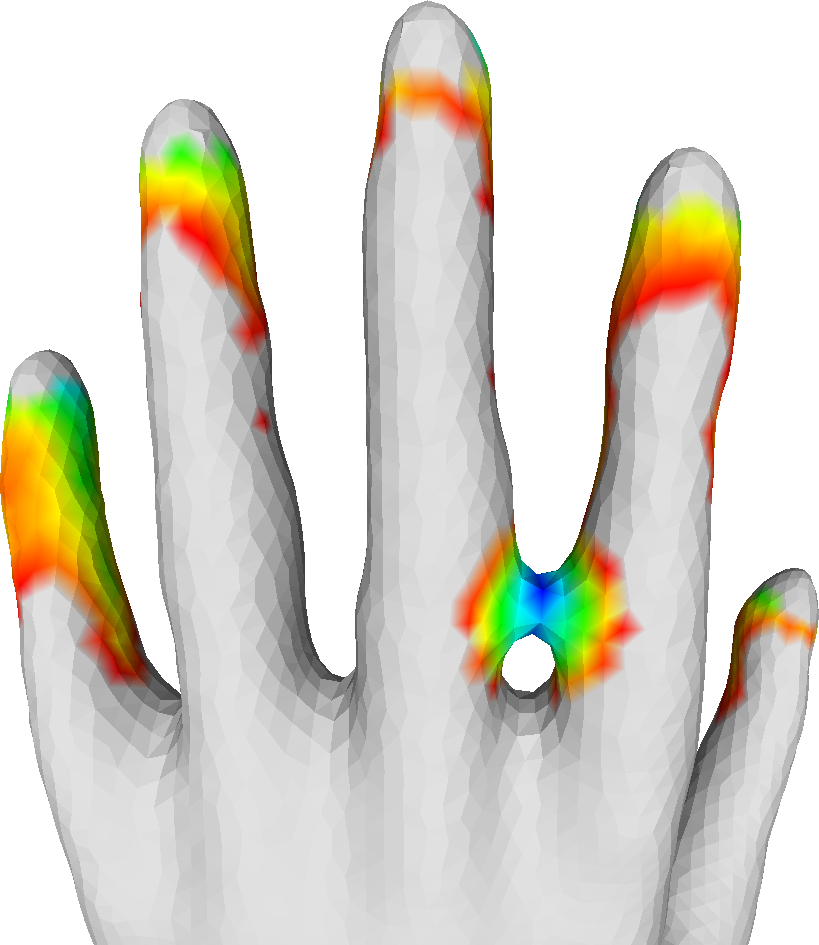}
	} \quad
	\subfigure[\label{fig:negative_all}]
	{
		\includegraphics[width=0.1\textwidth, angle=90]{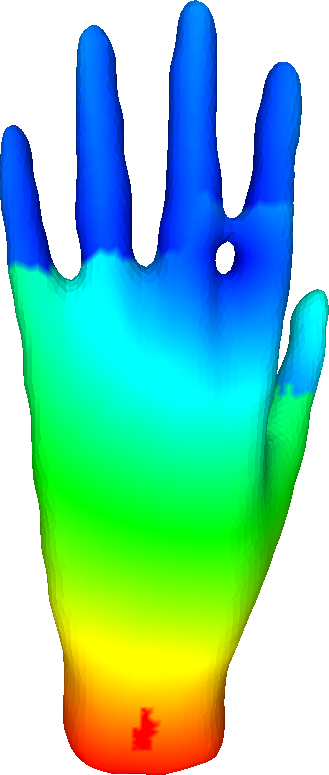}
	} \quad
	\subfigure[\label{fig:negative_thin}]
	{
		\includegraphics[width=0.1\textwidth]{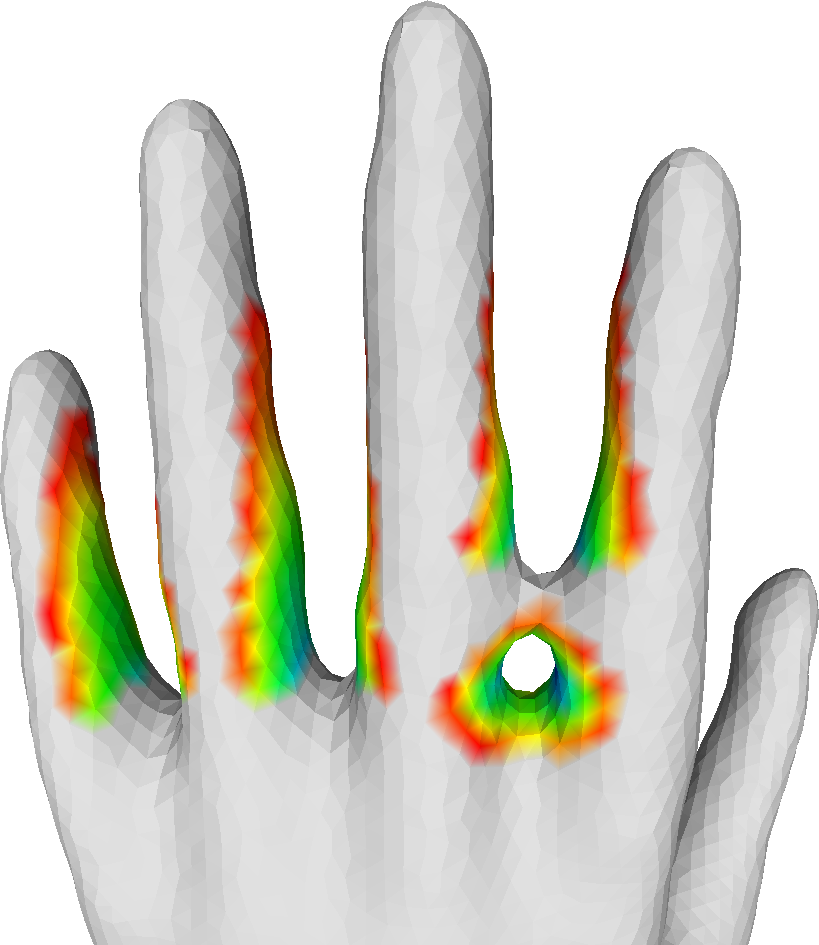}
	} \quad
	
	\caption{
		Detecting thin features in Model 187. Colors go from blue (thinnest parts) to red (thickest parts). \subref{fig:positive_all} Complete positive $\epsilon$-map. \subref{fig:negative_all} Complete negative $\epsilon$-map. \subref{fig:positive_thin} Partial positive $\epsilon$-map. \subref{fig:negative_thin} Partial negative $\epsilon$-map.
	}
	\label{fig:positive_negative_results}
\end{figure}

Figure \ref{fig:maps} depicts two complete and bi-directional $\epsilon$-maps, whereas Figures \ref{fig:positive_all} and \ref{fig:negative_all} show a positive and a negative complete map respectively.  

\begin{figure}[!h]
	\centering
	\includegraphics[width=0.3\textwidth]{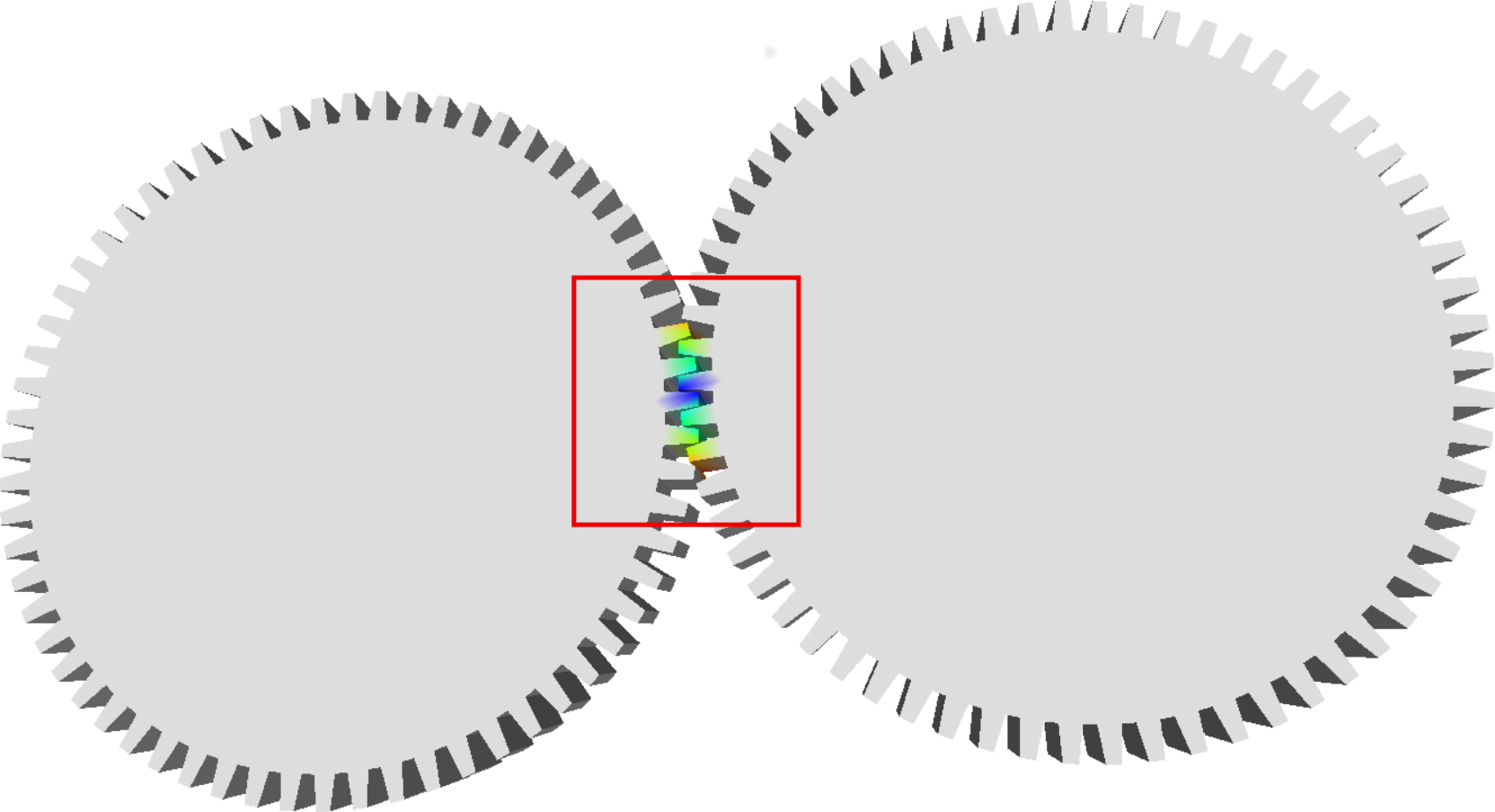} \quad
	\includegraphics[width=0.15\textwidth]{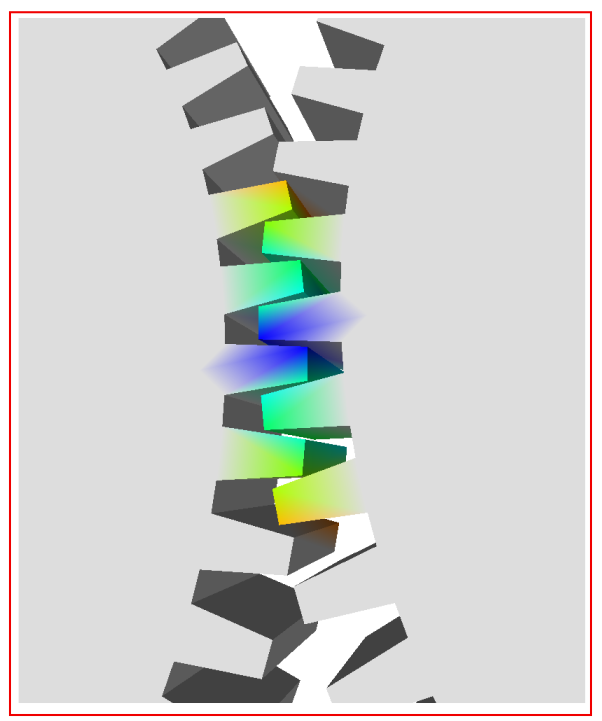}
	\caption{A negative partial $\epsilon$-map and a focus on detected thin features in the complementary part of the object.}
	\label{fig:gear}
\end{figure}

Figures \ref{fig:positive_thin} and \ref{fig:negative_thin} depict examples of positive and negative partial $\epsilon$-maps respectively. Note that the two partial maps highlight thin features detected in the model and in its complementary part respectively. As expected, the thin bridge connecting the two fingers on the right is detected as the thinnest feature in the object, while the cavity generated by the bridge and the gap between fingers is detected as thinnest features in its complementary part. An additional partial negative $\epsilon$-map is shown in Figure \ref{fig:gear}.

%
%
%

\begin{figure}[!h]
	\centering
	
	\subfigure[\label{fig:thick_1}
	] {\includegraphics[width=0.13\textwidth]{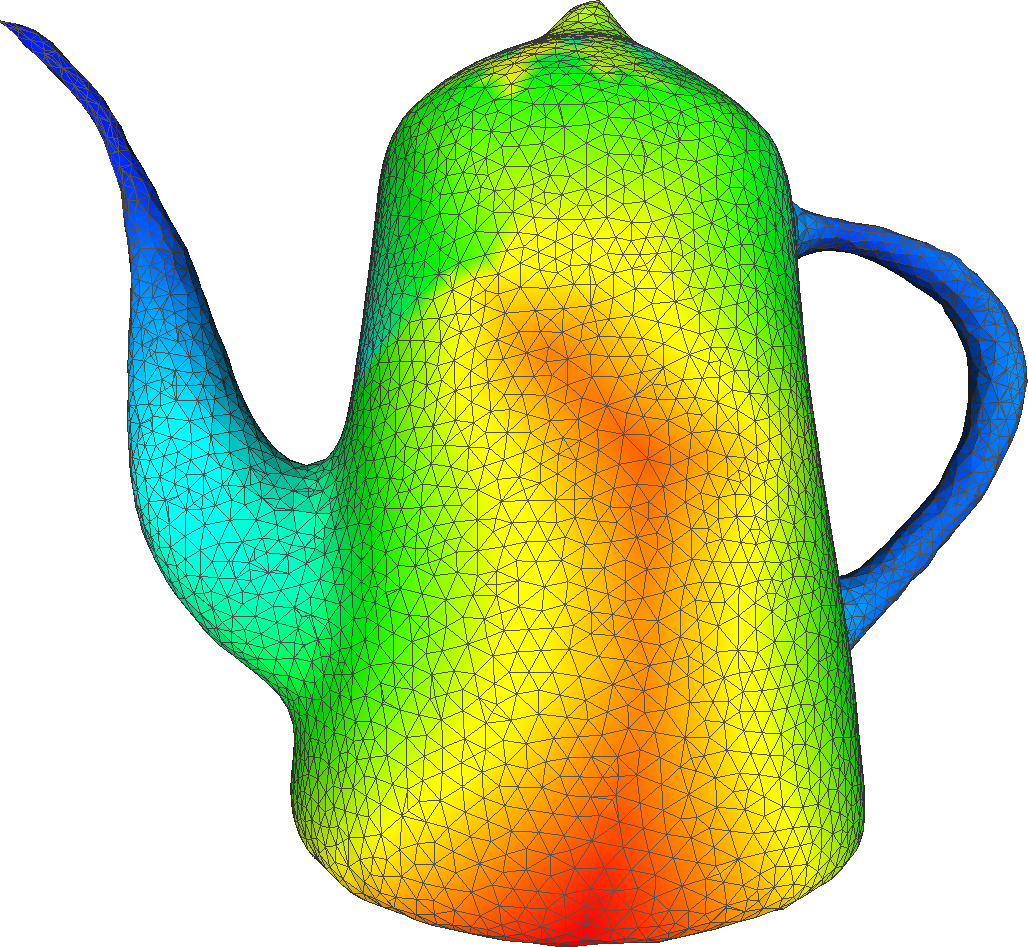} }
	\subfigure[\label{fig:thick_2}
	] 		{\includegraphics[width=0.13\textwidth]{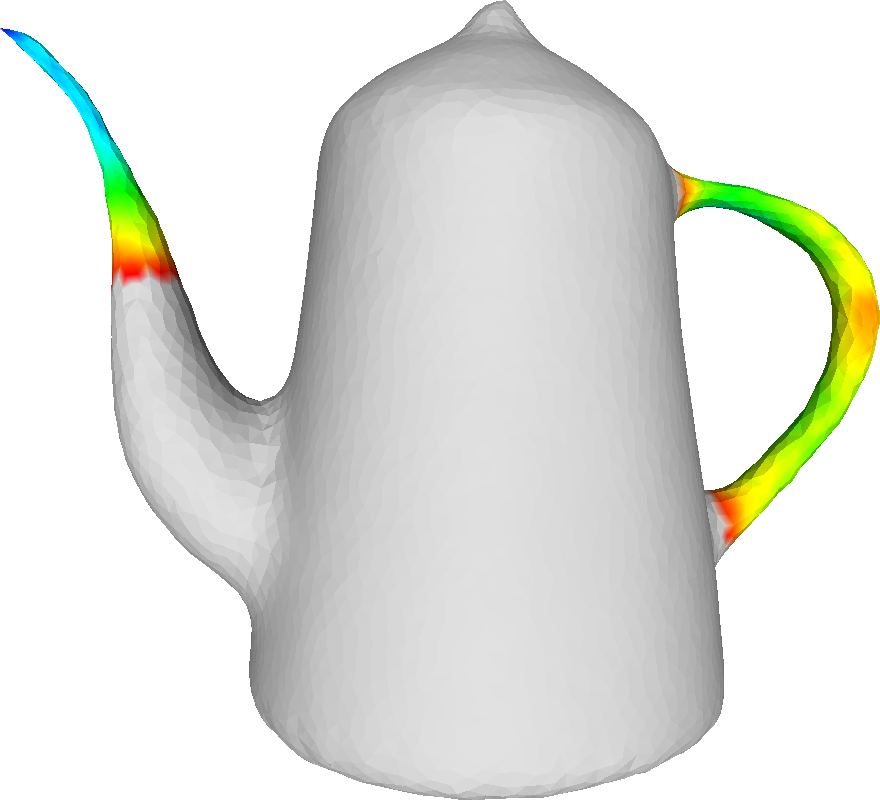} }
	\subfigure[\label{fig:thick_3}
	] 						{\includegraphics[width=0.13\textwidth]{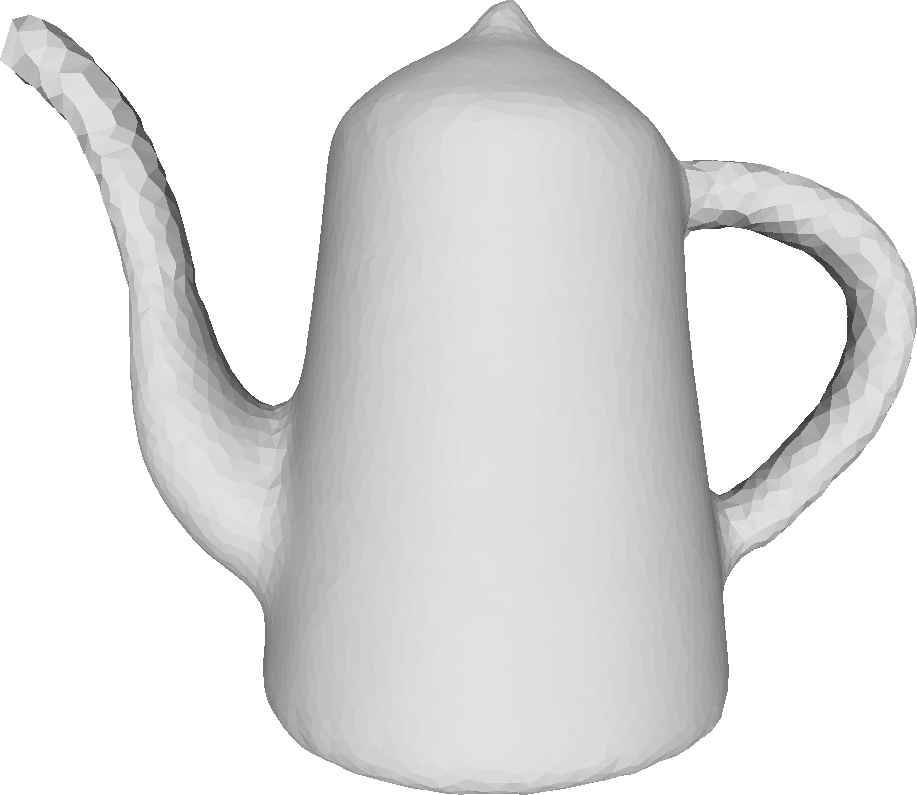} }
	
	
	\caption{
		Thickening. \subref{fig:thick_1} The complete positive $\epsilon$-map. \subref{fig:thick_2} The partial positive $\epsilon$-map. \subref{fig:thick_3}  The result of our thickening algorithm.
	}
	\label{fig:thickening_results}
\end{figure}

During the thickening, the result of the analysis is exploited to modify the input shape only where thickness is lower than the input threshold, while any other part is kept unchanged. Figure \ref{fig:thickening_results} shows both the complete and the partial positive $\epsilon$-maps for Model 372. Note that the latter is sufficient to run the thickening. In the example in Figure \ref{fig:thick_3}, only the handle and the spout of the teapot are edited as expected.

\begin{figure*}[!h]
	\centering
	
	\subfigure[\label{fig:thick124_0}]
	{
		\includegraphics[width=0.3\textwidth]{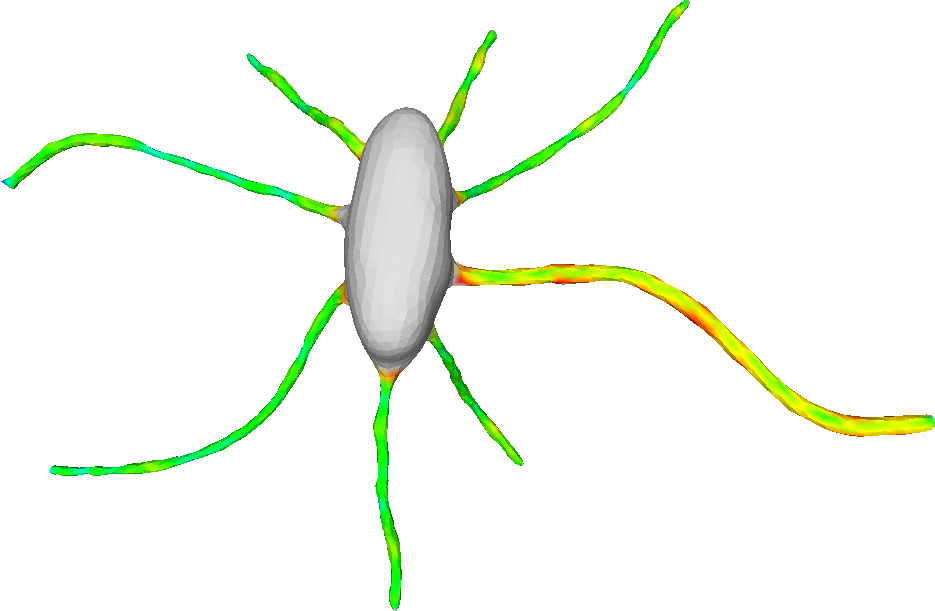} \quad
	}
	\subfigure[\label{fig:thick124_1}]
	{
		\includegraphics[width=0.3\textwidth]{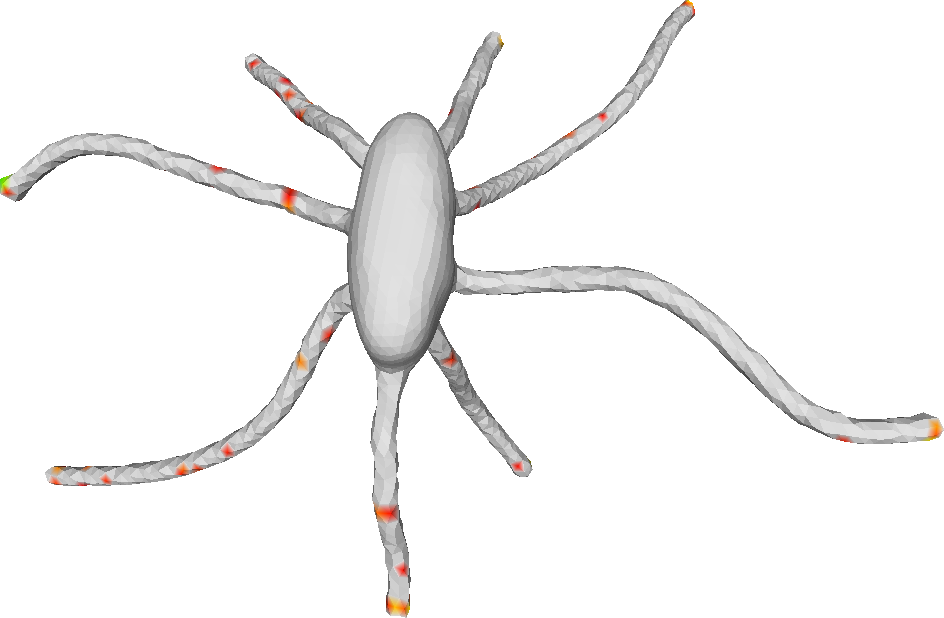} \quad
	}
	\subfigure[\label{fig:thick124_2}]
	{
		\includegraphics[width=0.3\textwidth]{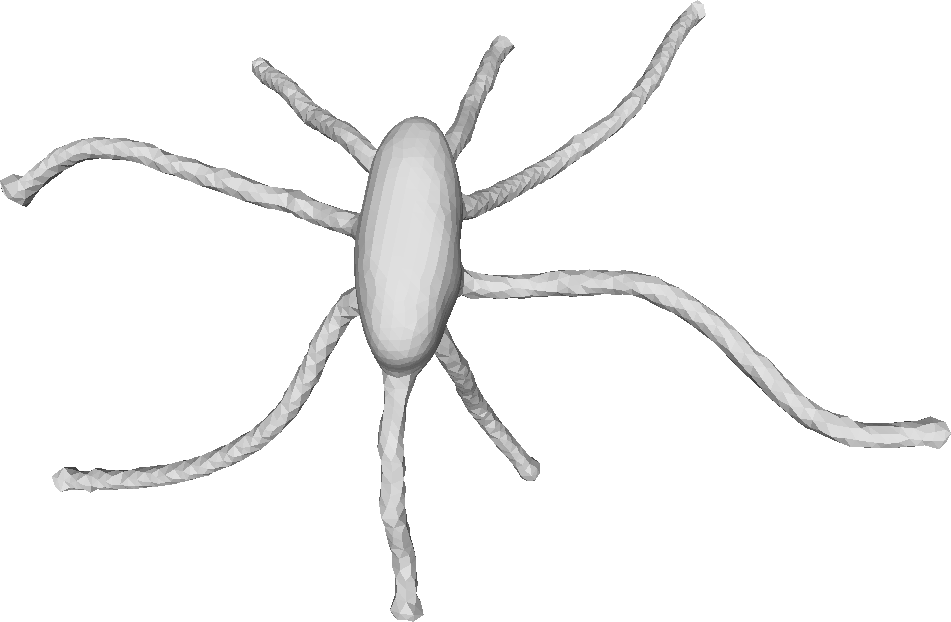} \quad
	}
	
	\caption{
		Thickening Model 124. Vertices mapped to a ``thick'' attribute are grey-colored. Thin vertices are underlined by colors: from blue (far from the thickness threshold) to red (closed to the threshold). \subref{fig:thick124_0} The partial positive $\epsilon$-map of the input shape. \subref{fig:thick124_1} The partial positive $\epsilon$-map after the application of our thickening algorithm (one iteration).  \subref{fig:thick124_1} The partial positive $\epsilon$-map of the final thickened object. 
	}
	\label{fig:thickened_examples}
\end{figure*}

\begin{figure*}[!h]
	\centering
	
	\subfigure[\label{fig:thick_99}
	] {
		\includegraphics[width=0.18\textwidth]{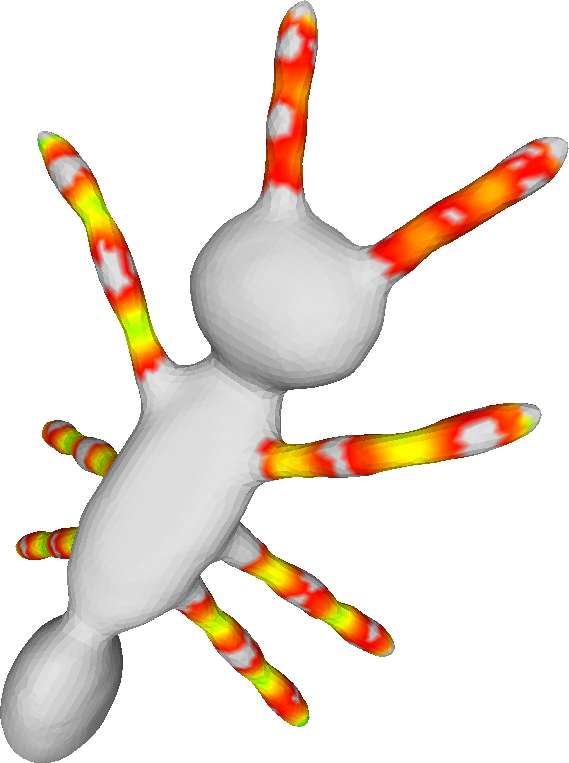}
		\includegraphics[width=0.18\textwidth]{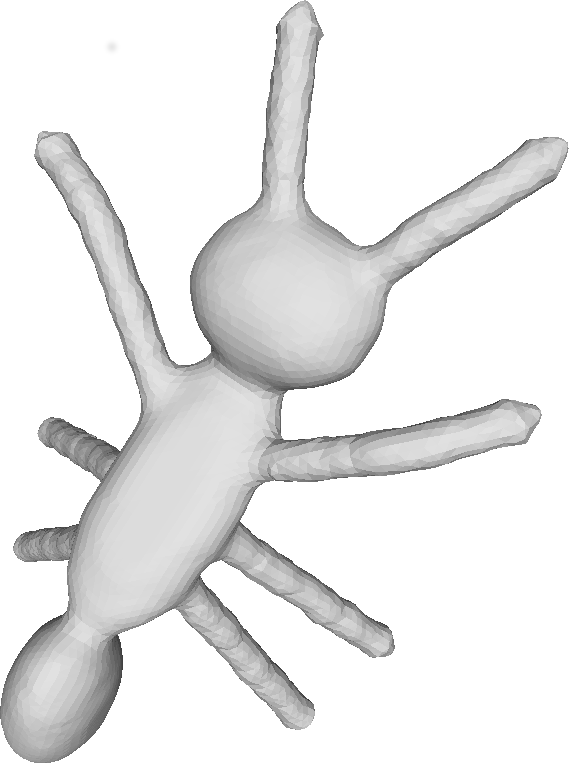} 
	}
	\qquad \quad
	\subfigure[\label{fig:thick_110}
	] {
		\includegraphics[width=0.15\textwidth]{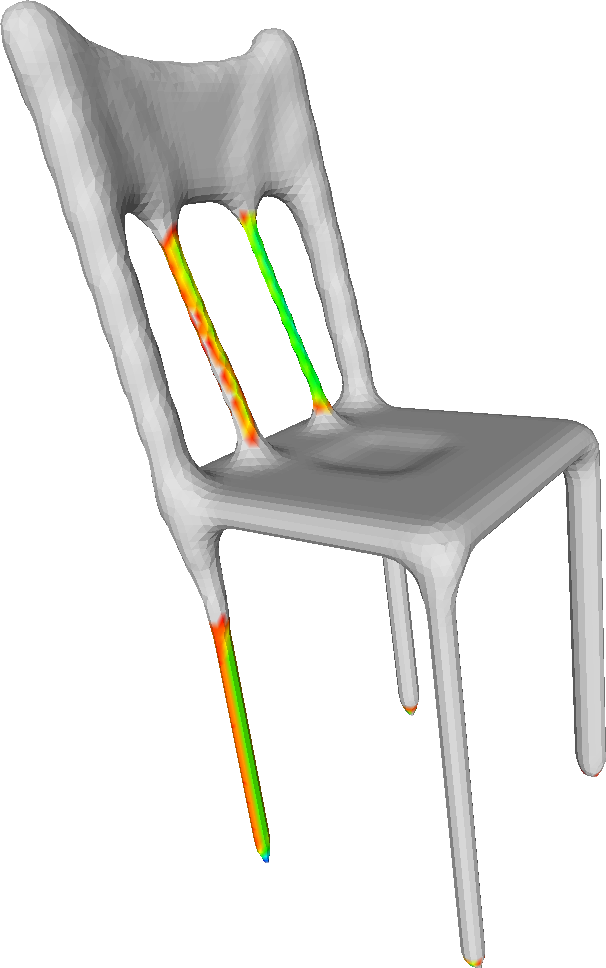} 
		\includegraphics[width=0.15\textwidth]{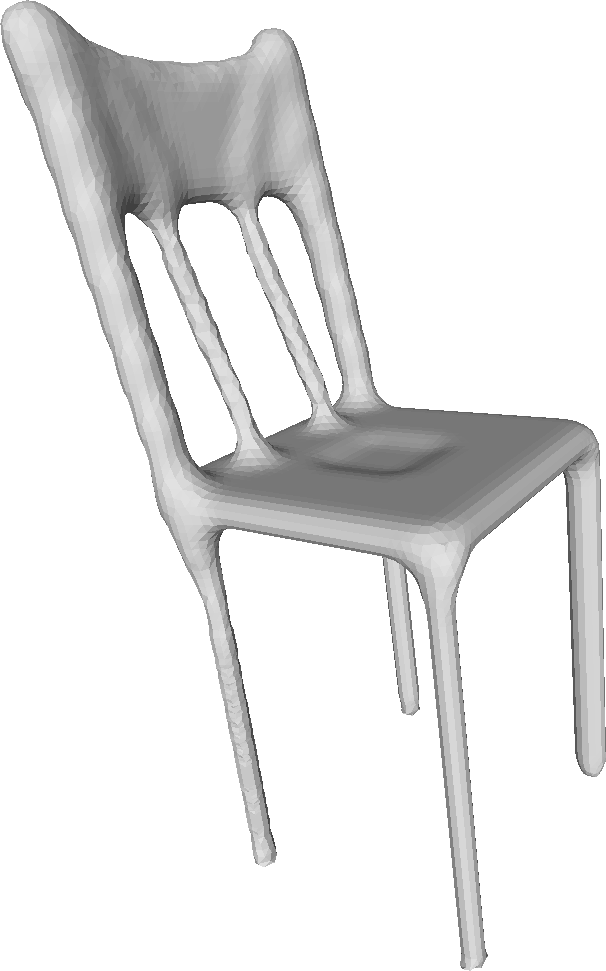} 
	} \qquad \quad
	
	\vspace{0.5cm}
	
	\subfigure[\label{fig:thick_129}
	] {
		\includegraphics[width=0.2\textwidth]{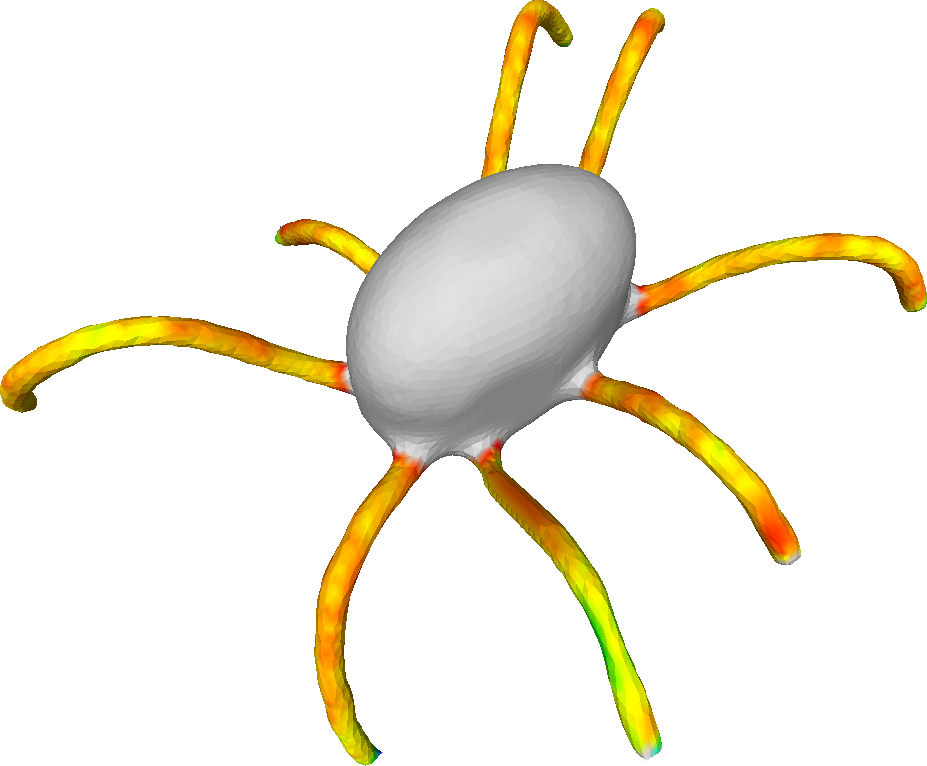} 
		\includegraphics[width=0.2\textwidth]{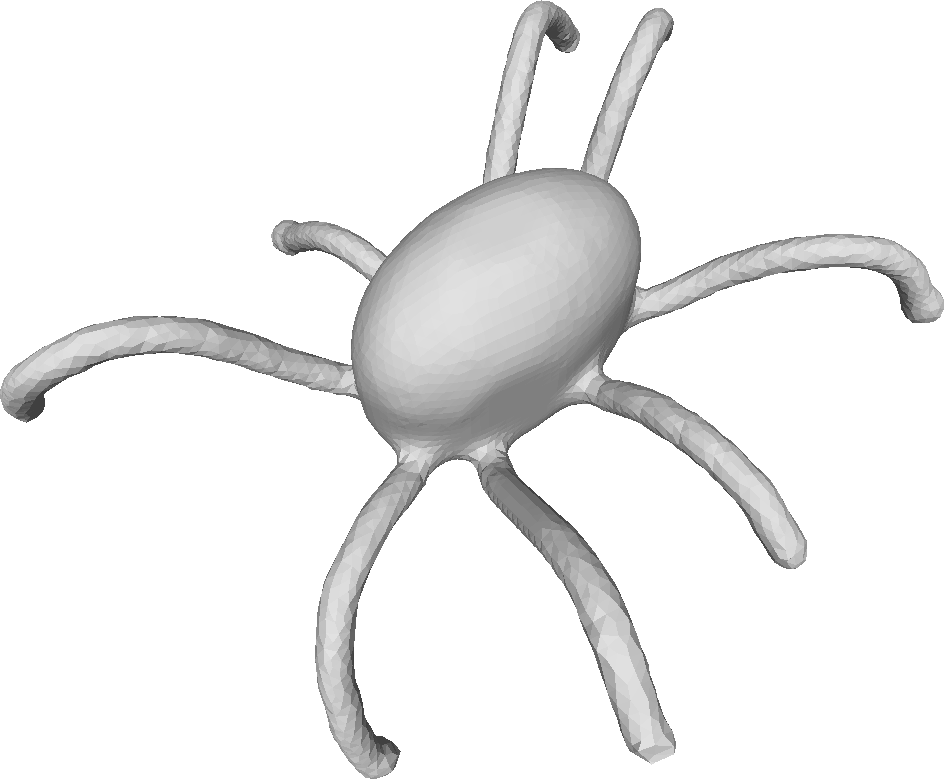} 
	} \qquad \quad
	\subfigure[\label{fig:thick_233}
	] {
		\includegraphics[width=0.18\textwidth]{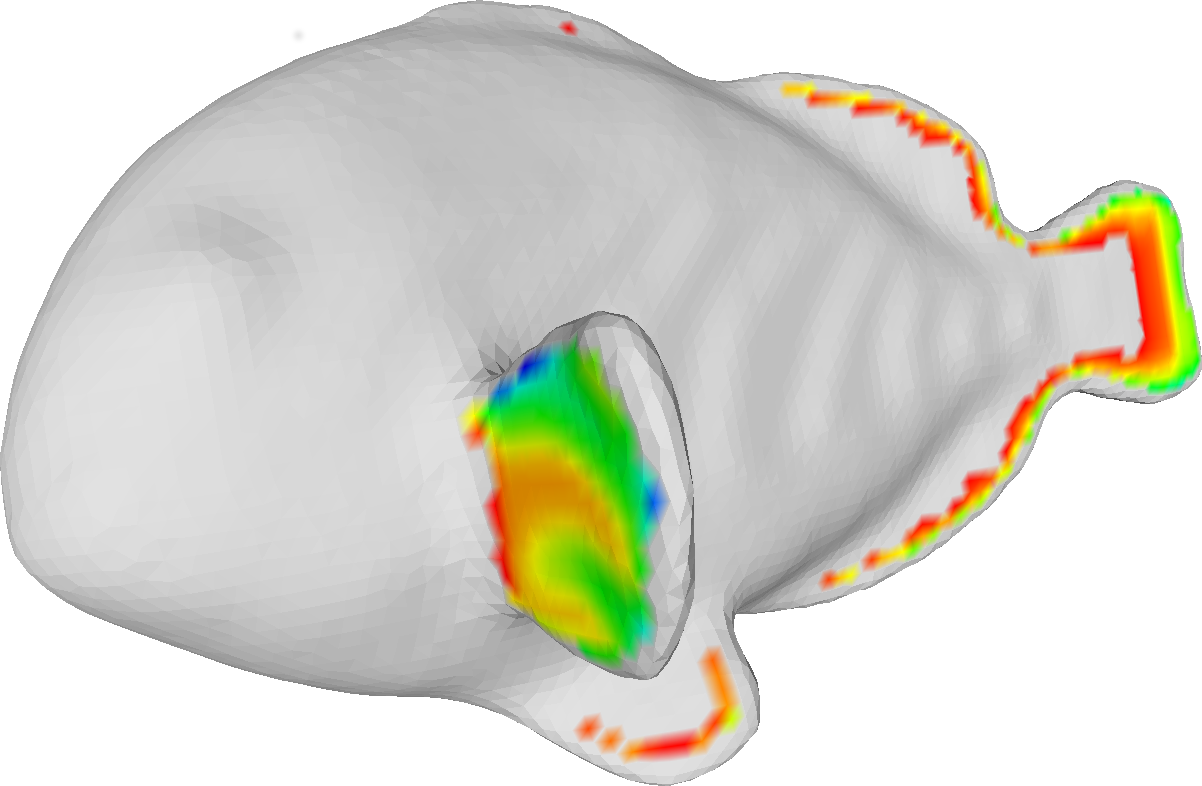} \quad
		\includegraphics[width=0.18\textwidth]{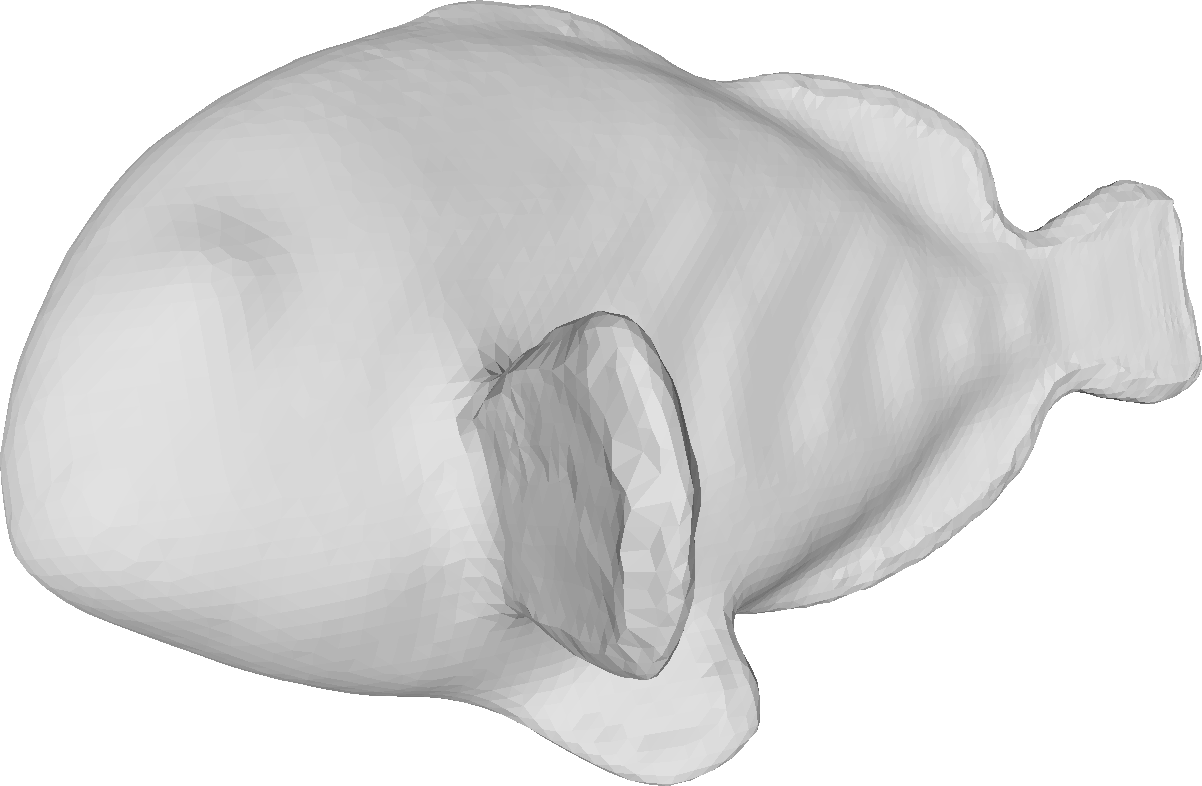} 
	} \qquad \quad
	
	\vspace{0.5cm}
		
	\subfigure[\label{fig:thick_45}
	] {
		\includegraphics[width=0.2\textwidth]{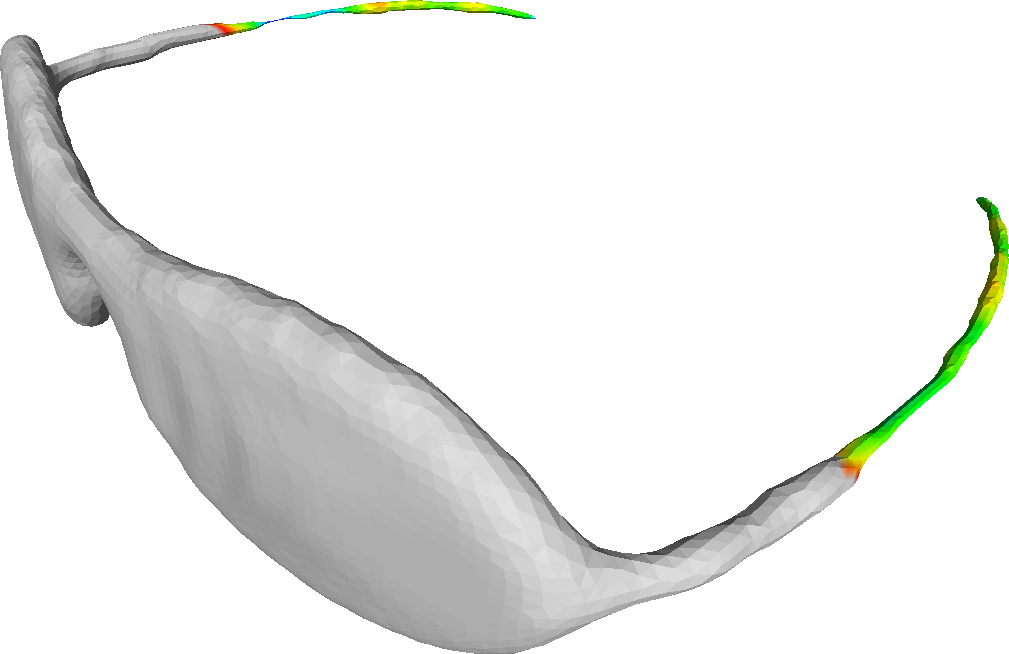} \quad
		\includegraphics[width=0.2\textwidth]{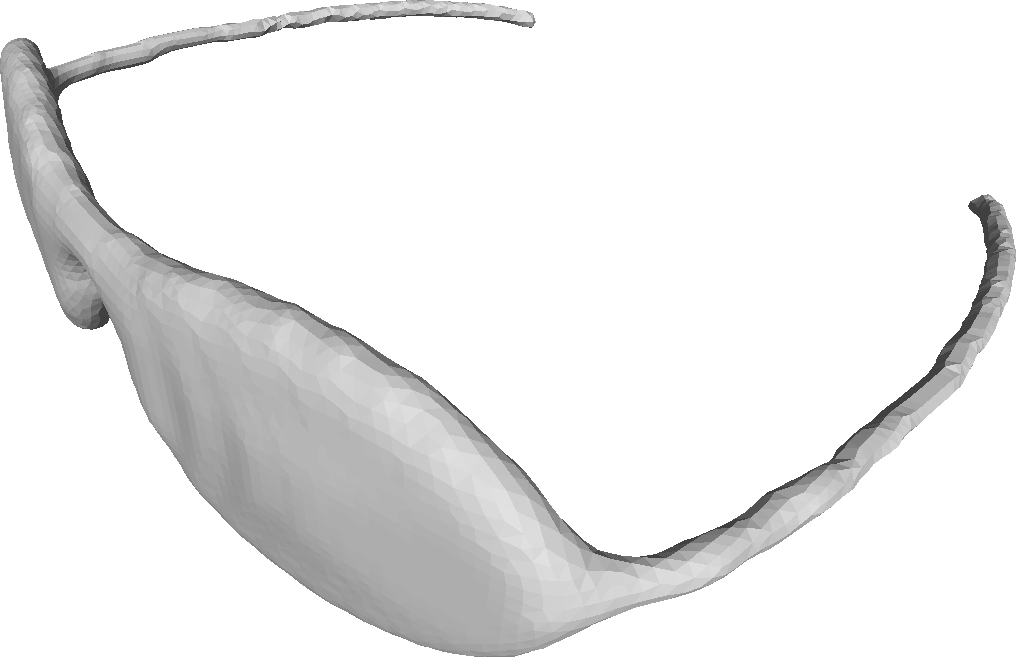} 
	} \qquad \quad
	\subfigure[\label{fig:thick_stiff}
	] {
		\includegraphics[width=0.16\textwidth]{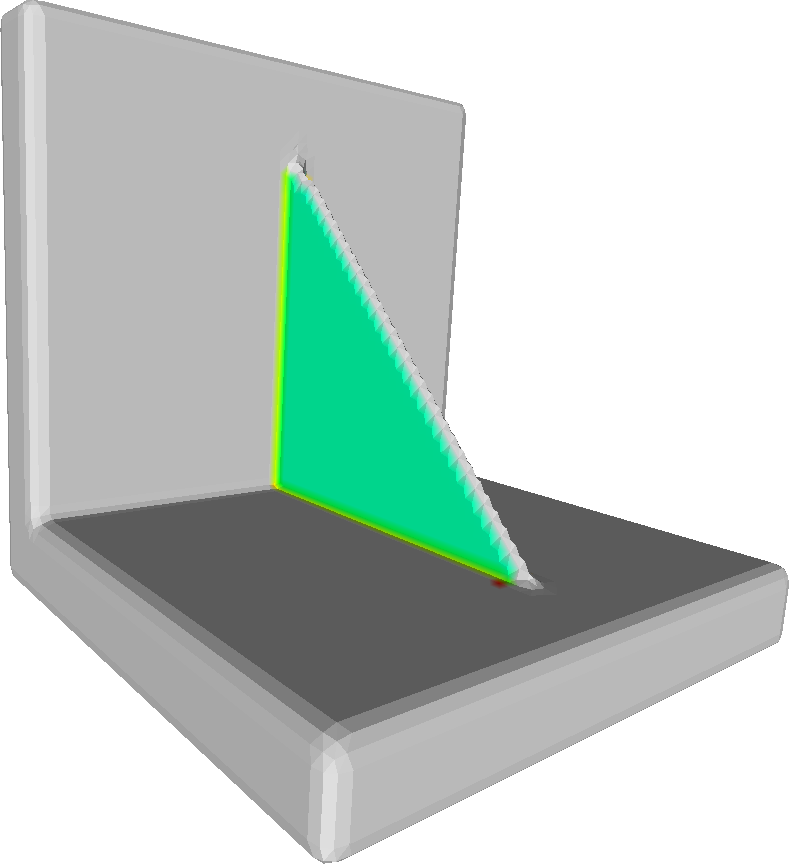} 
		\includegraphics[width=0.16\textwidth]{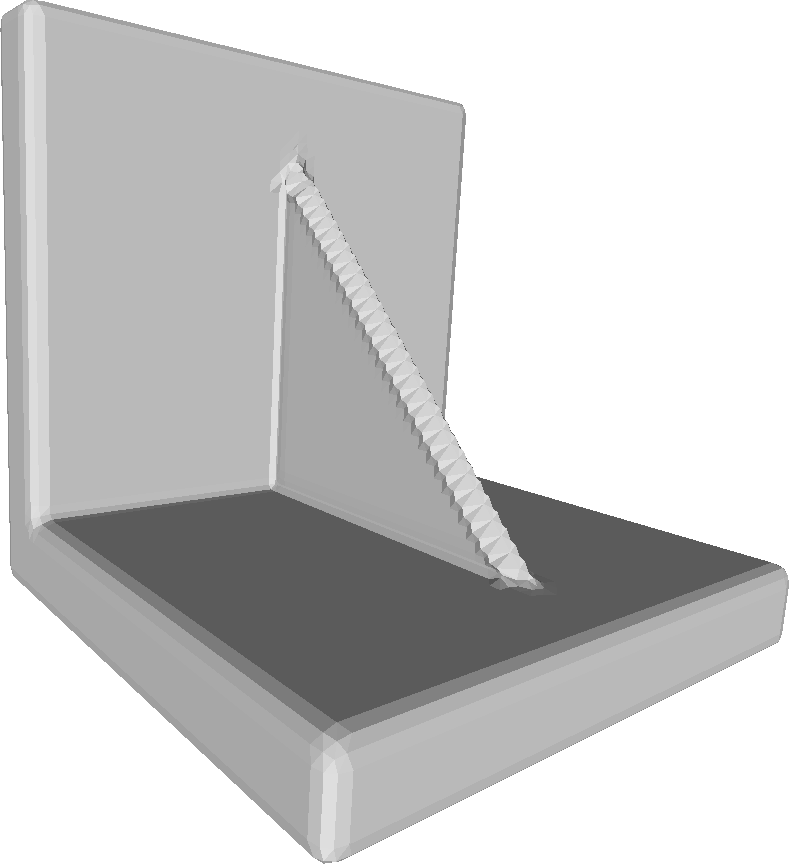} 
	}
	
	\caption{
		Results. For each mesh, a partial positive $\epsilon$-map is shown on the left, while the result of our thickening algorithm is shown on the right. Vertices mapped to a ``thick'' attribute are grey-colored. Thin vertices are underlined by colors: from blue (far from the thickness threshold) to red (closed to the threshold).
	}
	\label{fig:thickened_results}
\end{figure*}

In Figures \ref{fig:thickened_examples}, two positive $\epsilon$-maps are shown, which represent thickness information before and after the application of our thickening algorithm. Our heuristic approach actually increases the minimum thickness of the shape, but in some cases it is not sufficient to achieve the desired thickness threshold in one single iteration. In these cases, the whole algorithm can be run again, and we could verify that this process converged to an actual (positive) $\epsilon$-shape is all of our experiments. Figure \ref{fig:thickened_results} show some additional results.

\paragraph{Execution time} Experiments were run on a Linux machine (1.9GHz Intel Core i7, 16GB RAM). Table \ref{tab:times} shows the execution times referred to the computation of complete and partial $\epsilon$-maps. As expected, our algorithm is sensibly accelerated when a thickness threshold is specified.

At a theoretical level, the time needed to characterize a single vertex grows as the number of tetrahedra to be visited increases. The worst case is represented by a convex polyhedron whose $n$ vertices are all on the surface of a sphere. In this case, for each vertex $O(n)$ tetrahedra must be visited, which means that the overall complexity is $O(n^2)$. In contrast, if a shape is very thin everywhere and the tetrahedral mesh has no unnecessary internal Steiner points
, then the antipodean is normally found in the first few iterations. The use of a threshold to compute a partial map simulates this behaviour.
These theoretical considerations are confirmed by our experiments. As an example, the computational times referred to Model 192 (i.e. the hand) and Model 372 (i.e. the teapot) are sensibly different, even if both models have about 13.5K triangles.

Table \ref{tab:times} reports the time spent by our algorithm to compute the complete bi-directional $\epsilon$-map for a set of test models. These values were measured while running a sequential version of the algorithm, where simplexes are analyzed one after the other. However, it should be considered that our method is embarrassing parallel, that is, several simplexes of the same degree can be simultaneously analyzed. This is true for both the computation of complete and partial $\epsilon$-maps. Our implementation considers this aspect and exploits OpenMP \cite{Dagum:1988} to enable such a parallelism. Thanks to this parallel implementation, we succeeded in reducing the running time by a factor of 2.5 on our 4-core machine.

\begin{table}[!ht]
	
	\centering
	\small
	
	\begin{tabular}{|l|r|r|r|c|}
		\hline
		\multirow{2}{*}{\textbf{Input}} 	& \textbf{Input} 		& \textbf{Complete} 	& \textbf{Partial}   	& \textbf{Model }	\\
		& \textbf{Triangles} 	& \textbf{$\epsilon$-map} 	& \textbf{$\epsilon$-map } & \textbf{shown in}	\\
		\hline
		Model 400	& 7736		& 2720	 		& 45 		& Figure \ref{fig:sdf_result} \\		
		Model 121	& 12282		& 1995	  		& 405 		& Figure \ref{fig:map121} \\
		Model 372	& 13376		& 15380			& 320		& Figure \ref{fig:thickening_results} \\
		Model 192	& 13978		& 6295			& 430		& Figure \ref{fig:sdf_result} \\
		Model 197	& 15102		& 8250	  		& 422 		& Figure \ref{fig:map197} \\		
		Model 397 	& 15242		& 10640	 		& 420		& Figure \ref{fig:sdf_result} \\	
		Model 187	& 15678		& 13400		 	& 487 		& Figure \ref{fig:maps} \\		
		Model 130	& 15998		& 6605	  		& 586 		& Figure \ref{fig:sdf_result} \\

		\hline		
	\end{tabular}
	
	\caption{Execution times in seconds. Partial $\epsilon$-maps are computed by setting the thickness threshold to $1\%$ of the bounding box diagonal of the input mesh.}
	\label{tab:times}
	
\end{table}

\subsection{Comparison}
\begin{figure}[!h]
	\centering
	
	\includegraphics[width=0.48\textwidth]{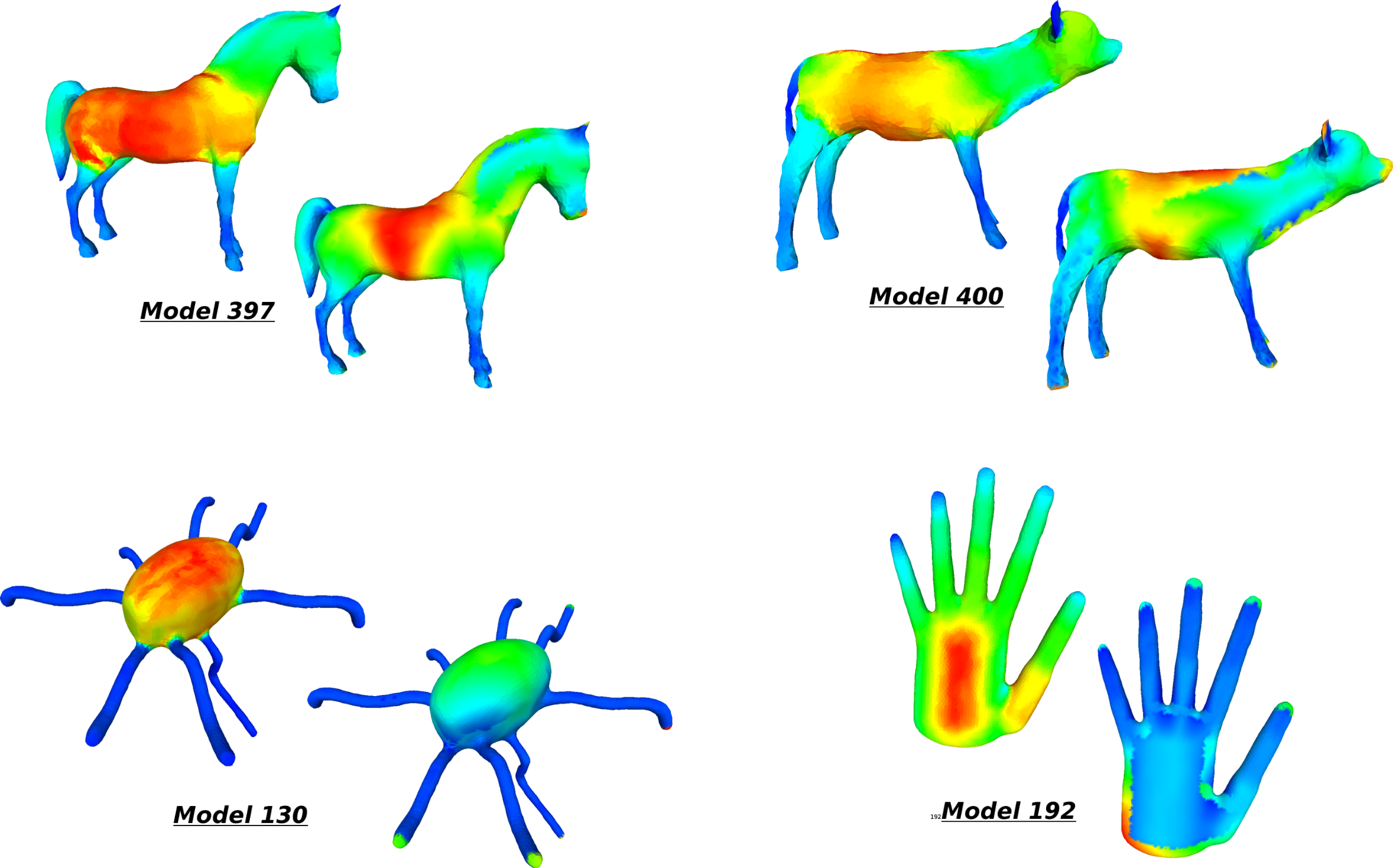}
	
	\caption{Comparison between SDF and $\epsilon$-shapes. For each input mesh, SDF is on the left and the complete $\epsilon$-map is on the right.}
	\label{fig:sdf_result}
\end{figure}

The Shape Diameter Function (SDF) can be considered the state-of-the-art tool for thickness evaluation on meshes. Its implementation available in CGAL \cite{cgal_sdf} has been exploited to analyze our dataset and compare the results. Figure \ref{fig:sdf_result} shows a comparison between complete $\epsilon$-maps and SDF. As expected, in some cases the two thickness-based descriptors provide a similar representation of the input shape.

\begin{figure}[!h]
	\centering
	
	\subfigure[SDF]
	{
		\includegraphics[width=0.17\textwidth]{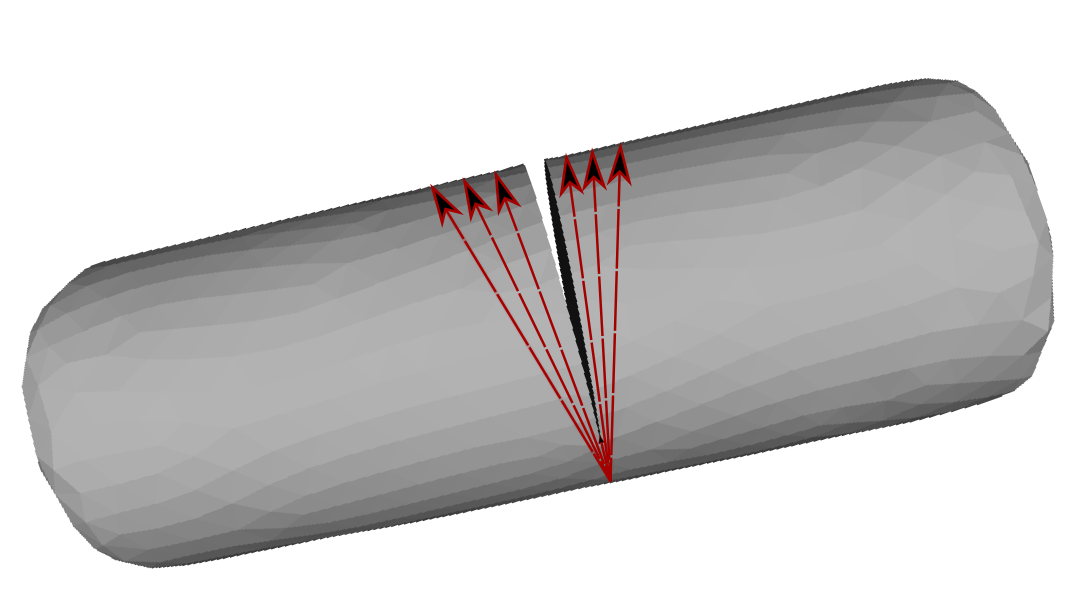} \qquad
		\includegraphics[width=0.22\textwidth]{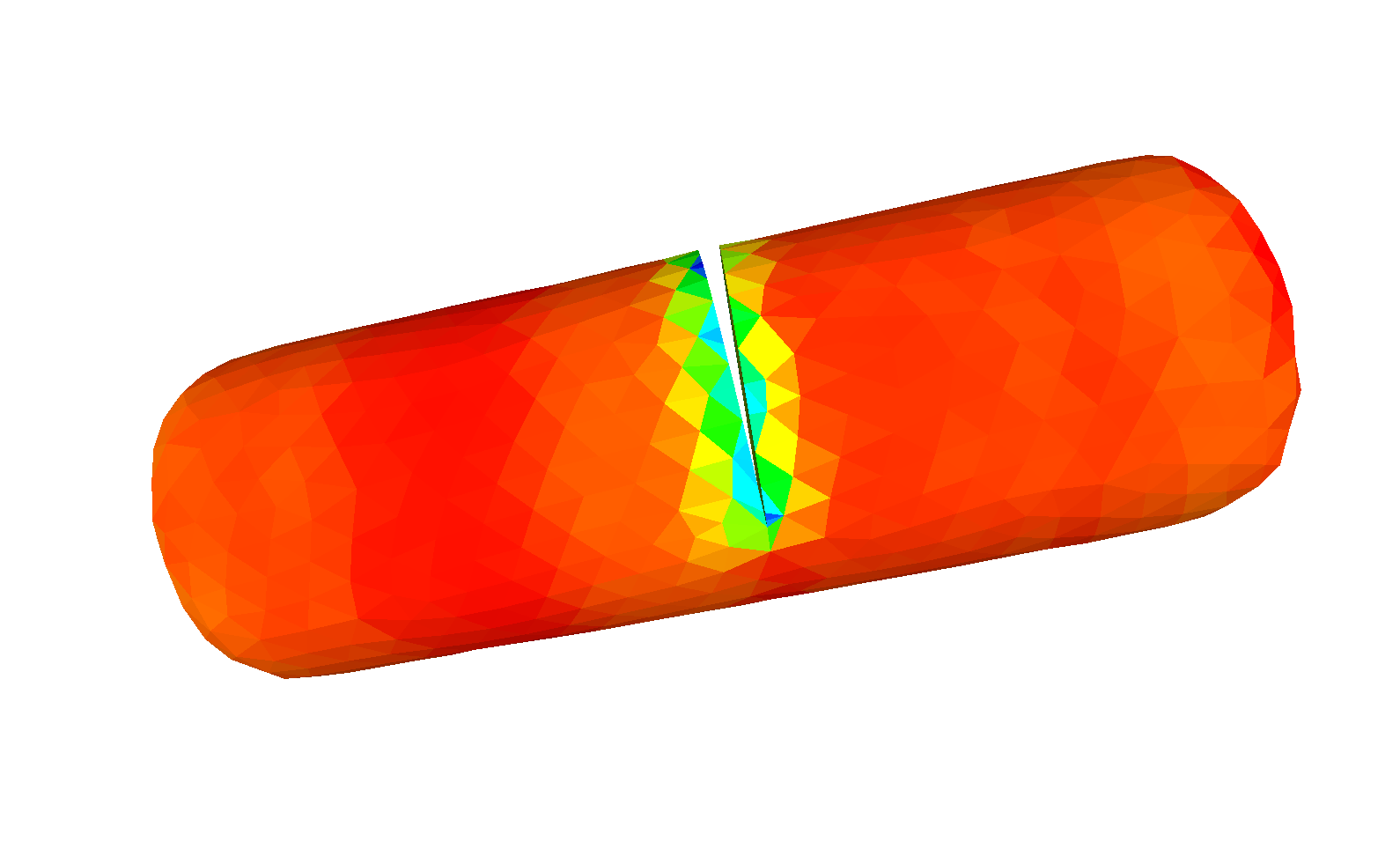}
	}
	\subfigure[Ours]
	{
		\includegraphics[width=0.17\textwidth]{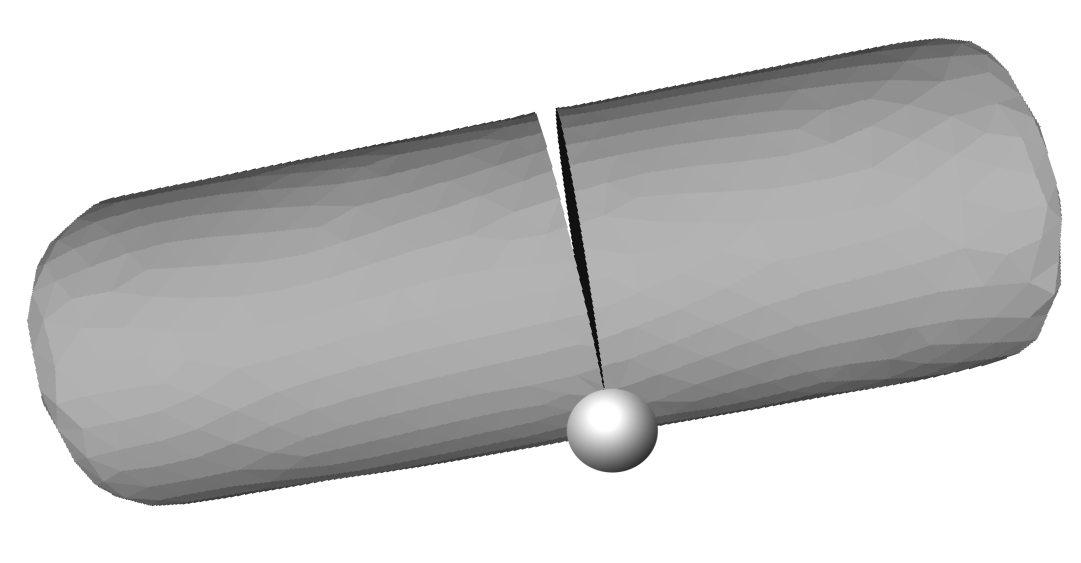} \qquad
		\includegraphics[width=0.22\textwidth]{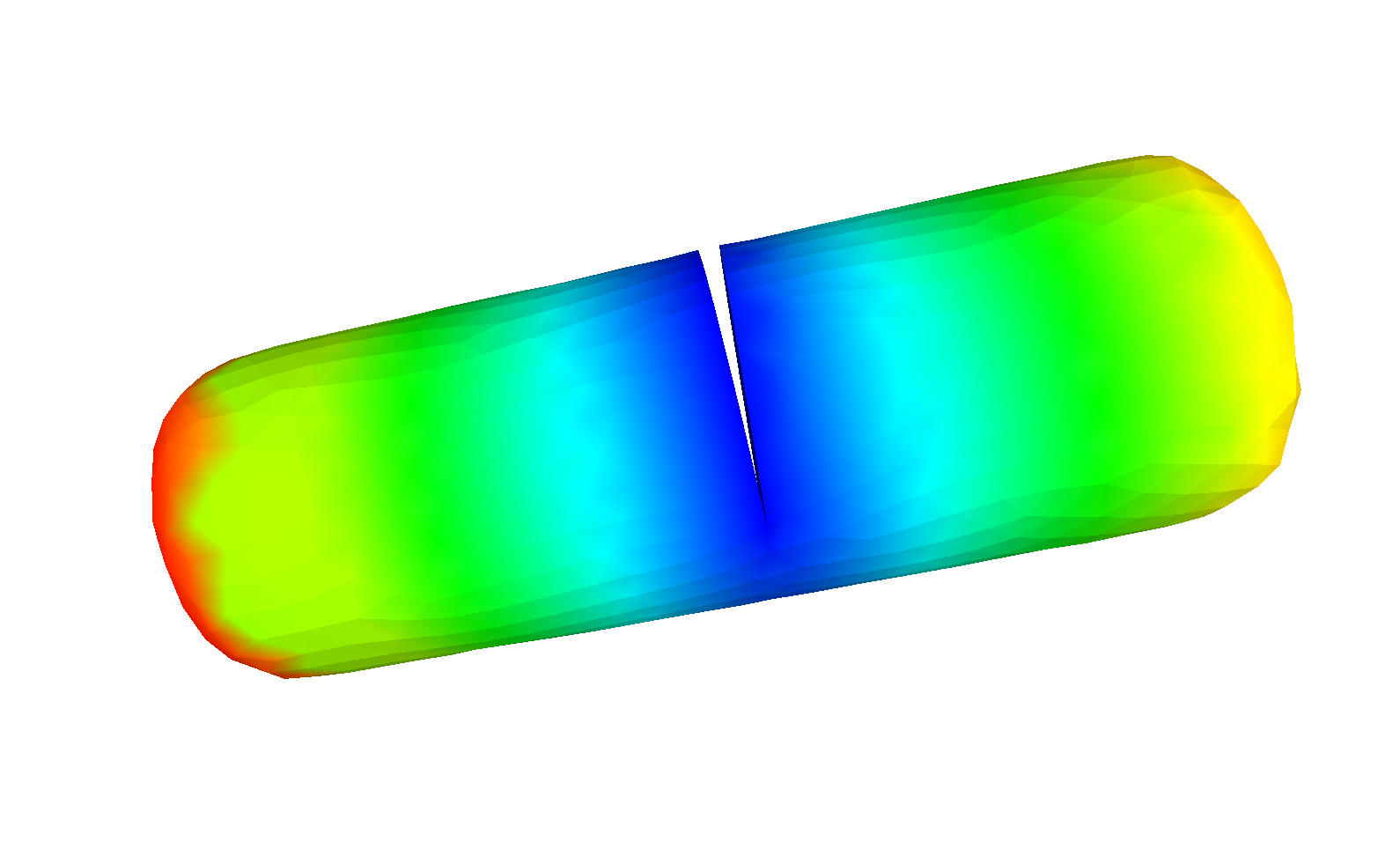}
	}
	
	\caption{Comparison between SDF and $\epsilon$-shapes. The statistic average of the distances applied by the SDF returns a high thickness value even at the bottom of the cylinder under the cut. Our method evaluates this area as the thinnest area of the object.}
	\label{fig:sdf_critical_results}
\end{figure}

Nevertheless, some relevant differences are visible. Consider thickness evaluation at the extremities of the shape (e.g. the fingertips of the hand in Figure \ref{fig:sdf_result}) and at some points closest to very narrow features (Figure \ref{fig:sdf_critical_results}). In both cases, the statistic average of the distances applied by the SDF provides an unexpected result. In the former case, most of the rays casted from an extremal point touch the surface closest to the point itself. The opposite happens when a narrow bottleneck is present. By increasing the number of casted rays a more precise result may be achieved but, though this would have a significant impact on the performances, no guarantee can be given in any case.

Our method solves this limitation and computes the exact thickness value even at these points according to our definition. Note that this exact value is guaranteed to be calculated and does not depend on any user-defined input parameter. Also, through our method narrow features can be efficiently detected by means of a partial $\epsilon$-map.

%
%

\section{Conclusions and Future Work}
\label{sec:future}

We have shown that both 2D polygons and 3D polyhedra can be automatically characterized based on their local thickness. Our novel mathematical formulation allows to implement effective algorithms that associate an exact thickness value to any relevant point of the object, without the need to set any unintuitive parameter, and without the need to rely on estimates or approximations.
Furthermore, we have shown how our analysis can be used to thicken thin features, so that models which are inappropriate for certain applications become suitable thanks to an automatic local editing.

While the analysis is rigorously defined, our thickening algorithm is still based on heuristics and cannot guarantee a successful result in all the cases. Stated differently, if $\epsilon$ is the threshold thickness used to perform the thickening, we would like to guarantee that the thickened model is an $\epsilon$-shape. Therefore, an interesting direction for future research is represented by the study of thickening methods that provide such a guarantee, and it is easy to see that such methods must necessarily be free to change the local topology.

\section*{Acknowledgment}
This project has received funding from the European Unions Horizon 2020 research and innovation programme under grant agreement No 680448 (CAxMan). Thanks are due to all the members of the Shapes and Semantics Modeling Group at IMATI-CNR and in particular to Silvia Biasotti for helpful discussions.

\section*{References}

\bibliographystyle{model3-num-names}
\bibliography{paper.bib}

\end{document}